\documentclass[bibyear]{aa}

\usepackage{hyperref}
\usepackage{graphicx}
\usepackage{amsmath}
\usepackage{amssymb,bm}
\usepackage{tabularx}
\usepackage{longtable}
\usepackage[table]{xcolor}
\usepackage{booktabs}
\usepackage[normalem]{ulem}
\usepackage{natbib}
\usepackage{float}
\usepackage{color}
\makeatletter
\renewcommand*\aa@pageof{, page \thepage{} of~\pageref*{LastPage}}
\hypersetup{colorlinks=true,allcolors=[rgb]{0,0,0.7}}

\begin{document}


\newcommand{\RefChange}[1]{{\textcolor{black}{#1}}}
\newcommand{\smallspace}{\vspace{-.35cm}}
\newcommand*\diff{\mathop{}\!\mathrm{d}}


\newcommand{\eagle}{\textsc{eagle}}
\newcommand{\auriga}{\textsc{auriga}}
\newcommand{\agama}{\textsc{agama}}
\newcommand{\emosaics}{\textsc{emosaic}s}
\newcommand{\Au}{\textsc{Au}}
\newcommand{\MW}{\textsc{MW}}
\newcommand{\Gaia}{\textit{Gaia}}
\newcommand{\LCMD}{\ensuremath{\Lambda\mathrm{CDM}}}

\newcommand{\GroupFont}[1]{{\fontfamily{ptm}\selectfont{\textsl{#1}}}}
\newcommand{\Disc}{\GroupFont{Disc}}
\newcommand{\Halo}{\GroupFont{Halo}}
\newcommand{\Ungrouped}{\GroupFont{Ungrouped Halo}}
\newcommand{\GES}{\GroupFont{GES}}
\newcommand{\HTD}{\GroupFont{HTD}}
\newcommand{\HS}{\GroupFont{HS}}
\newcommand{\Sequoia}{\GroupFont{Sequoia}}
\newcommand{\LRLsixtyfour}{\GroupFont{LRL64}}
\newcommand{\EDone}{\GroupFont{ED-1}}
\newcommand{\EDtwo}{\GroupFont{ED-2}}
\newcommand{\EDthree}{\GroupFont{ED-3}}
\newcommand{\EDfour}{\GroupFont{ED-4}}
\newcommand{\EDfive}{\GroupFont{ED-5}}
\newcommand{\EDsix}{\GroupFont{ED-6}}
\newcommand{\Thamnos}{\GroupFont{Thamnos}}
\newcommand{\Typhon}{\GroupFont{Typhon}}

\newcommand{\kpc}{\ensuremath{~\text{kpc}}}
\newcommand{\Mpc}{\ensuremath{~\text{Mpc}}}
\newcommand{\Gpc}{\ensuremath{~\text{Gpc}}}
\newcommand{\MpcCube}{\ensuremath{\,\text{Mpc}^{-3}}}
\newcommand{\MpcInv}{\ensuremath{\,\text{Mpc}^{-1}}}
\newcommand{\Msun}{\ensuremath{\,\text{M}_\odot}}
\newcommand{\kms}{\ensuremath{\,\text{km}\hspace{0.1em}\text{s}^{-1}}}
\newcommand{\uJ}{\ensuremath{\,\mathrm{kpc}\,\mathrm{km}/\mathrm{s}}}
\newcommand{\uE}{\ensuremath{\,\mathrm{km}^2/\mathrm{s}^2}}
\newcommand{\uEe}{\ensuremath{10^{5}\,\mathrm{km}^2/\mathrm{s}^2}}

\newcommand{\J}{\ensuremath{\bf{J}}}
\newcommand{\sqJ}{\ensuremath{ \tilde{\boldsymbol{J}}}}
\newcommand{\FobsJ}{\ensuremath{F_{\mathrm{obs}}\left(\mathbf{J}\right)}}
\newcommand{\FobsJcorrect}{\ensuremath{F_{\mathrm{obs}}^{*}\left(\mathbf{J}\right)}}
\newcommand{\FJ}{\ensuremath{F\left(\mathbf{J}\right)}}
\newcommand{\FJs}{\ensuremath{F_{\odot}\left(\mathbf{J}\right)}}
\newcommand{\FJt}{\ensuremath{F\left(\boldsymbol{J},\boldsymbol{\theta}\right)}}
\newcommand{\FobsJt}{\ensuremath{F_{\mathrm{obs}}\left(\mathbf{J},\mathbf{\theta}\right)}}
\newcommand{\FobsJtcorrect}{\ensuremath{F_{\mathrm{obs}}^{*}\left(\mathbf{J},\mathbf{J}\right)}}
\newcommand{\SJ}{\ensuremath{S_{\mathrm{Obs}}\left(\mathbf{J}\right)}}

\newcommand{\Jr}{\ensuremath{J_{\textnormal{r}}}}
\newcommand{\JR}{\ensuremath{J_{\textnormal{R}}}}
\newcommand{\Jz}{\ensuremath{J_{\textnormal{z}}}}
\newcommand{\Jtotal}{\ensuremath{J_{\textnormal{total}}}}
\newcommand{\Lz}{\ensuremath{L_{\textnormal{z}}}}
\newcommand{\Met}{\ensuremath{\left[\mathrm{Fe}/\mathrm{H}\right]}}
\newcommand{\Emin}{\ensuremath{E_{\min}}}
\newcommand{\wJ}{\ensuremath{w\left(\boldsymbol{J}\right)}}
\newcommand{\tJ}{\ensuremath{\tau\left(\boldsymbol{J}\right)}}
\newcommand{\vtoomre}{\ensuremath{V_{\mathrm{Toomre}}}}
\newcommand{\vtan}{\ensuremath{V_{\mathrm{Tan}}}}
\newcommand{\vmax}{\ensuremath{V_{\max}}}
\newcommand{\vmed}{\ensuremath{V_{\mathrm{med}}}}
\newcommand{\pmin}{\ensuremath{p_{\min}}}

\newcommand{\FiJ}{\ensuremath{F_{i}\left(\mathbf{J}\right)}}
\newcommand{\FX}{\ensuremath{F\left(\mathbf{X}\right)}}
\newcommand{\X}{\ensuremath{\boldsymbol{X}}}

\newcommand{\Rsun}{\ensuremath{R_{0}}} 
\newcommand{\Msol}{\ensuremath{M_{\odot}}}
\newcommand{\vlos}{\ensuremath{v_{\mathrm{los}}}}
\newcommand{\ud}{\ensuremath{\boldsymbol{u}^{5D}}}

\newcommand{\Act}{\ensuremath{ \left( \boldsymbol{J} , \boldsymbol{\theta} \right) }}
\newcommand{\Cart}{\ensuremath{ \left( \boldsymbol{x} , \boldsymbol{v} \right) }}
\newcommand{\sPDF}{\ensuremath{\sigma_{\mathrm{PDF}}}}

\newcommand{\NhsT}{\ensuremath{N^{\mathrm{HS}}_{\mathrm{Total}}}}
\newcommand{\NhsU}{\ensuremath{N^{\mathrm{HS}}_{\resizebox{0.1cm}{0.1cm}{{$\uparrow$}}}}}
\newcommand{\NhsD}{\ensuremath{N^{\mathrm{HS}}_{\resizebox{0.1cm}{0.1cm}{{$\downarrow$}}}}}

\defcitealias{dodd23GaiaDR3View}{Dodd23}
\defcitealias{mikkola23NewStellarVelocity}{Mikkola23}
\defcitealias{naik24MissingRadialVelocities}{Naik24}


\title{Filling in the blanks:}
\subtitle{A method to infer the substructure membership and dynamics of 5D stars}

\author{
Thomas M. Callingham
\and
Amina Helmi
}

\institute{
Kapteyn Astronomical Institute, University of Groningen, Landeleven 12,
9747 AD Groningen, The Netherlands\\
\email{t.m.callingham@astro.rug.nl}
}

\date{Accepted XXX. Received YYY; in original form ZZZ}


\abstract{ In the solar neighbourhood, only $\sim2\%$ of stars in the \Gaia{} survey have a line-of-sight velocity (\vlos) contained within the RVS catalogue.
 These limitations restrict conventional dynamical analysis, such as finding and studying substructures in the stellar halo.
 }
 { \RefChange{We aim} to present and test a method to infer a probability density function (PDF) for the missing \vlos{}
 of a star with 5D information within $2.5\,\mathrm{kpc}$.
 This technique also allows us to infer the probability that a 5D star is associated with the Milky Way's
 stellar Disc or the stellar Halo, which can be further decomposed into known stellar substructures.
 }
 { We use stars from the \Gaia{} DR3 RVS catalogue to describe the local orbital structure in action space.
 The method is tested on a 6D \Gaia{} DR3 RVS sample and a 6D \Gaia{} sample crossmatched to ground-based spectroscopic surveys,
 stripped of their true \vlos. 
 The stars predicted \vlos, membership probabilities, and inferred structure properties
 are then compared to the true 6D equivalents,
 allowing the method's accuracy and limitations to be studied in detail.
 }
 {Our predicted \vlos{} PDFs are statistically consistent with the true \vlos{}, 
 with accurate uncertainties.
 We find that the \vlos{} of Disc stars can be well-constrained, with a median uncertainty of $26\,\kms$. 
 Halo stars are typically less well-constrained with a median uncertainty of $72\,\kms$,
 but those found likely to belong to Halo substructures can be better constrained.
 The dynamical properties of the total sample and subgroups,
 such as distributions of integrals of motion and velocities, are also accurately recovered.
 The group membership probabilities are statistically consistent with our initial labelling,
 allowing high-quality sets to be selected from 5D samples by choosing a trade-off between
 higher expected purity and decreasing expected completeness.
 }
 {
 We have developed a method to estimate 5D stars' \vlos{} and substructure membership.
 We have demonstrated that it is possible to find likely substructure members
 and statistically infer the group's dynamical properties.
 }

\keywords{ Galaxy: Halo \textendash{} galaxies: kinematics and dynamics \textendash{} methods: numerical
\textendash{} techniques: radial velocities }

\maketitle



\section{Introduction}\label{Sec:Intro}
The Milky Way (MW) has grown over its lifetime by consuming many smaller galaxies,
adhering to the hierarchical growth characteristic of the \LCMD{} cosmological model \citep{davis85EvolutionLargescaleStructure}.
Evidence of this assembly history can still be found recorded in the MW's halo,
which contains the stellar material of these cannibalised galaxies 
\citep[e.g.][]{helmi99DebrisStreamsSolar,bullock05TracingGalaxyFormation,cooper10GalacticStellarHaloes}.
Once identified, these accreted stars can be studied by
combining dynamical information with analysis of their stellar population,
such as chemistry and age. 
Together, these clues allow us to characterise the properties of destroyed satellites
and uncover a picture of our Galaxy's past.

However, disentangling stars into their progenitor groups can be challenging,
particularly for older accretion events.
The most recent accretion events can be seen in physical space and on the sky as streams,
such as the Sagittarius dwarf galaxy \citep{ibata94DwarfSatelliteGalaxy}.
Over time, the stars of accreted galaxies spread out over their orbits, 
becoming extended, diffuse stellar distributions
that are much harder to identify, even more so for larger progenitors \citep{helmi99DebrisStreamsSolar,helmi03PhasespaceStructureCold}.
Fortunately, most of these stars remain on orbits similar to their progenitor,
preserving substructure in integrals of motion (IoM) space \citep[e.g.][]{gomez10IdentificationMergerDebris}.
Many previous works have tried to exploit this dynamical coherence to identify substructures
in the hopes of inferring the assembly history of our Galaxy 
\citep{naidu20EvidenceH3Survey,callingham22ChemoDynamicalGroupsGalactic}.

Typically, dynamically detecting individual substructures requires large samples of stars with accurate kinematics
\citep{helmi00MappingSubstructureGalactic,helmi20StreamsSubstructuresEarly}.
The \Gaia{} mission \citep{gaiacollaboration16GaiaMission}
has  convincingly delivered such data,
substantially enriching our understanding of the MW and its assembly history.
Following the release of \Gaia{} DR2 \citep{gaia-collaboration18GaiaDataRelease},
it was confirmed that the inner stellar halo is dominated by a single massive accreted dwarf galaxy,
\Gaia{} Enceladus Sausage (GES) \citep{belokurov18CoformationDiscStellar,helmi18MergerThatLed}.
Soon after, further halo substructure was found in the \Gaia{} dataset 
such as Sequoia \citep{myeong19EvidenceTwoEarly},
Thamnos \citep{koppelman19MultipleRetrogradeSubstructures}, among others 
\citep{yuan18StarGONewMethod,myeong22MilkyWayEccentric}.
The most recent \Gaia{} data release, DR3 \citep{gaiacollaboration23GaiaDataRelease},
has been found to contain even more substructure.
Works such as~\cite{dodd23GaiaDR3View}, hereafter \citetalias{dodd23GaiaDR3View},
identify new dynamical groups labelled ED-1 to ED-6,
alongside other recent discoveries such as Typhon \citep{tenachi22TyphonPolarStream}.

Arguably, the driving force behind \Gaia's substantial success in identifying past accretion events has been the size of its 6D dataset;
stars with complete 6D position-velocity information are needed for dynamical analysis.
The essential 5D coordinates provided by \Gaia's astrometry (parallax, sky position, and proper motions)
are completed with a line-of-sight velocity (\vlos)
contained within the Radial Velocity Spectrometer (RVS) catalogue \citep{cropper18GaiaDataRelease}.
\Gaia{} DR2's RVS catalogue contained an impressive $7.2$ million stars.
\Gaia{} DR3's RVS catalogue grew to an astonishing $33.8$ million \citep{katz23GaiaDataRelease}.

However, the current RVS sample has a much lower limiting magnitude of $G<14$ (13 for DR2) compared to
a limit of $G<21$ for the astrometry.
As a result, the 5D sample is approximately 50 times larger,
containing around 1.4 billion stars with positions and proper motions.
Furthermore, even within local distances of a few kpc, the RVS sample is biased towards intrinsically brighter stars.
Most of the stars in a local sample are part of the MW's stellar disc,
which is approximately a few hundred to a thousand times larger than the
stellar halo in which the accreted substructure resides.
Once a halo selection is made, combined with additional quality and distance cuts,
the RVS DR3 sample is severely reduced;
the local $2.5\,\mathrm{kpc}$ halo sample used in \citetalias{dodd23GaiaDR3View} contained approximately $69$ thousand stars.

These limitations associated with working within the RVS catalogue could be avoided if the lack of \vlos{} 
could be overcome or mitigated.
Some works have identified likely halo stars in the 5D dataset using cuts in transverse sky velocity
\citep{gaiacollaboration18GaiaDataRelease_HR}, or defined samples based on reduced proper motions 
\citep{koppelman21ReducedProperMotion,viswanathan23HiddenDeepHalo}.
These 5D halo datasets can then be used by techniques such as the {\verb|STREAMFINDER|} algorithm,
which can find coherent stellar streams upon the sky using 5D information \citep{ibata21ChartingGalacticAcceleration}.
For specific phase-mixed structures,
the effect of missing \vlos{} information can be minimised by focusing on specific regions of the sky, such as the Galactic anticentre.
In \citet{koppelman19CharacterizationHistoryHelmi} and \citet{ruiz-lara22UnveilingEvolutionProgenitor},
this approach is used to identify candidate members of the Helmi Streams \citep{helmi99DebrisStreamsSolar},
 which can be distinguished through $\mu_{b}$ alone because of their prominent vertical velocity.

An alternative approach is to attempt to constrain and estimate the missing \vlos.
Using the 6D \Gaia{} RVS sample,
statistical relationships between the 5D coordinates and the \vlos{} can be built,
allowing the likely \vlos{} of a star to be inferred.
In~\citet{mikkola23NewStellarVelocity} (hereafter \citetalias{mikkola23NewStellarVelocity}), 
a penalised maximum-likelihood method \citep{dehnen98DistributionNearbyStars,mikkola22VelocityDistributionWhite} is used to infer full velocity distributions
of the 5D \Gaia{} sample. 
This methodology reveals features in velocity space hidden in the 5D sample,
which can then be used to estimate the number of stars expected to belong to specific stellar substructures.
Similarly, \citet{naik24MissingRadialVelocities} (hereafter \citetalias{naik24MissingRadialVelocities})
is the latest in a series of papers \citep{dropulic21MachineLearningSixth,dropulic23RevealingMilkyWay}
that use  \RefChange{neural networks} trained on a 6D \Gaia{} RVS sample to predict the missing \vlos{} of a 5D \Gaia{} sample.
\RefChange{The work of  \citetalias{naik24MissingRadialVelocities},
uses Bayesian neural networks \citep{titterington04BayesianMethodsNeural} to infer}
predicted probability distributions with a typical uncertainty of $25$-$30$ \kms.


Both of these studies focus on the total distributions of velocities rather than attributing individual stars to specific halo substructures.
In principle, the predicted \vlos{} of individual stars could be used to complete the 6D position-velocity,
allowing IoM to be calculated and structure membership established.
To do this effectively and reliably,
the predicted \vlos{} must be well-constrained and accurate, especially for halo stars and members of substructures.
However, a common challenge of these approaches is that building the statistical relationship
between the entire 5D space and \vlos{} is difficult, particularly for members of the stellar halo.
Disc stars typically lie close to the plane and inhabit a reduced velocity range,
whereas halo stars can plausibly have a wide range of positions, velocities, and orbits.
Furthermore, fewer stars are distributed across the larger halo phase-space than that of the disc,
and stars of some smaller substructures can be spread extremely thinly.

In this paper, we develop and test a method to constrain the \vlos{} of stars with 5D information,
specifically targeting \RefChange{h}alo substructure.
We use the known density of a local sample of 6D \Gaia{} RVS stars in action-angle space
as a blueprint to determine how stars with 5D coordinates are likely to be dynamically distributed.
By exploiting the assumption that the majority of substructures are mostly phase-mixed,
with considered exceptions for those that are only partly phase-mixed,
action-angle space approximately reduces to action space.
This 3D space is significantly better populated by our 6D halo sample,
offering a considerable improvement over observable coordinates or cartesian space.
Furthermore, as IoM, an advantage of action space is that accreted groups can be more easily identified.
This trait allows us to infer the star's membership probability to the stellar disc, halo, or substructure therein.

The structure of the paper is as follows.
Section~\ref{Sec:Data} describes our 6D \Gaia{} RVS sample.
Section~\ref{Sec:ActionSpace} introduces action-angle space, details how we account for the selection effects of our sample,
and describes our methodology to evaluate the density in the space.
Section~\ref{Sec:Method} describes our method to calculate a \vlos{} PDF of a 5D star,
which can then be decomposed into the contribution from individual substructures
to find the probabilities of substructure membership.
In Section~\ref{Sec:PDFs}, we apply our methodology to the 6D sample and consider the statistical accuracy of 
the inferred PDFs, our ability to constrain the \vlos{} for different groups of stars, and marginalising over \vlos{} to 
recover dynamical distributions.
In Section~\ref{Sec:Probs}, we study our calculated membership probabilities and compare them to the true values,
exploring the trade-off between purity and completeness when selecting samples.
Section~\ref{Sec:Comparsion} compares our results to those of other works and techniques.
In Section~\ref{Sec:Discussion}, we discuss our results and methodology.
Finally, Section~\ref{Sec:Conclusion} summarises and concludes the paper.

\section{Data}\label{Sec:Data}
\begin{figure*}
 \includegraphics[width=\textwidth]{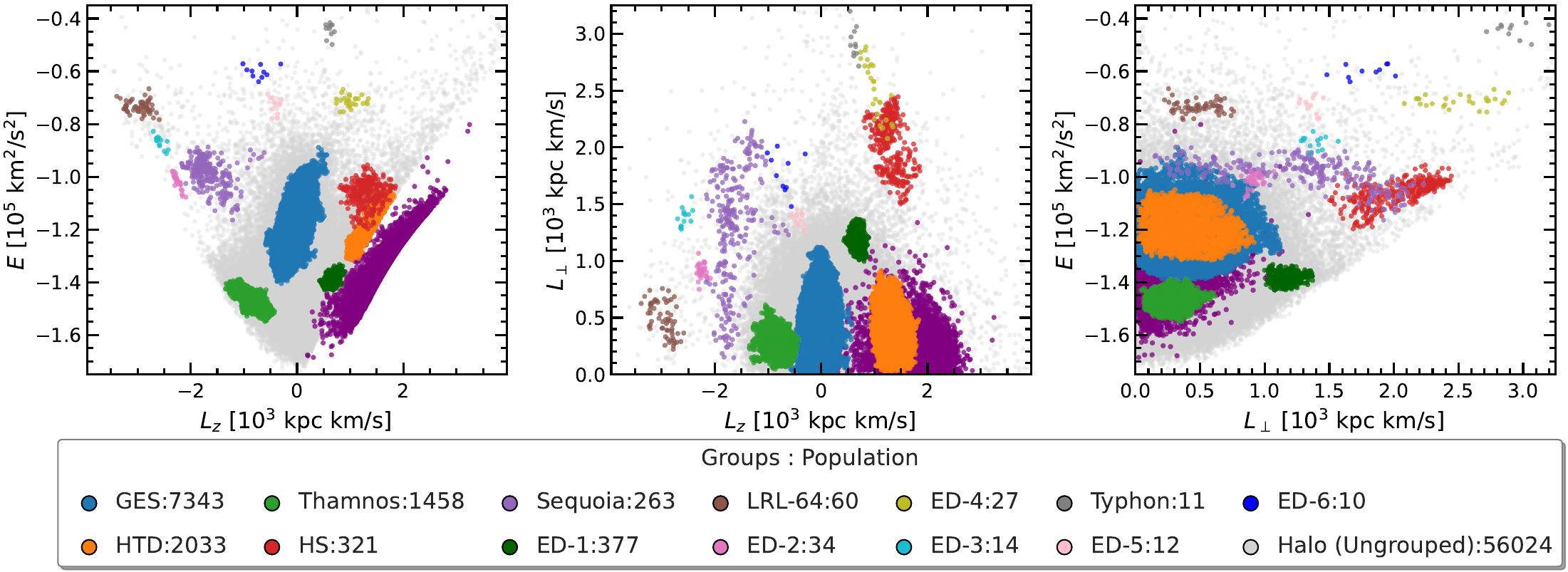}
 \vspace{-12pt}
 \caption{Local sample of Gaia DR3 stars in different dynamical spaces (panels),
 classified into different groups (indicated by colour) by \protect{\citetalias{dodd23GaiaDR3View}}.
 The panels show the spaces of energy and the vertical and perpendicular components of angular momentum
 used in \protect{\citetalias{dodd23GaiaDR3View}} to identify the \Halo{} groups.
 The \Disc{} group (purple) is kinematically selected with a Toomre velocity cut;
 the rest of the structures are found in the complementary kinematically selected \Halo.
 In the plot, scatter points representing the \Disc{} group have been down-sampled by a factor of 500,
 whilst the population in the legend is the total population of \Disc{} stars.
 \vspace{-10pt}
 }\label{fig:Data_Dodd23}
\end{figure*}

Our primary 6D dataset is based on the \Gaia{} DR3 RVS sample.
We apply several selection and quality cuts following \citet{dodd23GaiaDR3View}.
First, the parallaxes are corrected for their individual zero-point offsets \citep{lindegren21GaiaEarlyData}.
We require the relative (corrected) parallax uncertainty to be less than $20\%$,
a {\verb|RUWE|} value less than 1.4,
uncertainty on \vlos{} less than $20\kms$ (after applying~\cite{babusiaux23GaiaDataRelease} correction),
and $\left(G_{\mathrm{RVS}}-G\right)<-3$
\citep[also following][]{babusiaux23GaiaDataRelease}.

Distance is calculated by inverting the parallax.
We select stars within $2.5\,\mathrm{kpc}$, ensuring that parallax inversion is valid
and stars sufficiently populate the phase-space to estimate the density reliably.
For stars at low latitude ($|b| < 7.5^{\circ}$), we require higher signal-to-noise ratios for the spectra 
(\verb|rv_expected_sig_to_noise|~$>5$) following \citet{katz23GaiaDataRelease}.
We remove stars associated with GCs, as listed in \citet{baumgardt21AccurateDistancesGalactic}.
We remove stars that are found to be unbound (energy greater than $0$)
using the MW potential of \citet{cautun20MilkyWayTotal}.
This final sample contains 17,183,913 stars.

To convert to Galactocentric coordinates, we assume a
solar motion of ${\left(U, V, W\right)}_\odot = (11.1, 12.24, 7.25) \,\kms$ \citep{schonrich10LocalKinematicsLocal},
a local standard of rest of $\left|\textbf{V}_{\textrm{LSR}}\right|=232.8\kms$ \citep{mcmillan17MassDistributionGravitational},
and Solar radius of $\RefChange{\Rsun}=8.2\,\mathrm{kpc}$ \citep{mcmillan17MassDistributionGravitational}. 
We use the potential of~\cite{cautun20MilkyWayTotal},
which differs from \citetalias{dodd23GaiaDR3View},
but we find no substantial effect on the groupings of \citetalias{dodd23GaiaDR3View}.
For dynamical calculations, we use the software package \textsc{agama} \citep{vasiliev19AGAMAActionbasedGalaxy},
which uses the St\"{a}ckel fudge algorithm to calculate actions \citep{binney12ActionsAxisymmetricPotentials,sanders16ReviewActionEstimation}.

This data sample is divided kinematically into \Halo{} and \Disc{} stars
\footnote{Using {\GroupFont{a font}} to denote kinematically defined groups}
 using a Toomre velocity cut, 
where \Halo{} stars must have:
\begin{equation}
 V_\textrm{Toomre}=\left|\textbf{V}-\textbf{V}_\textrm{LSR}\right|> 210\,\kms. 
\end{equation}
This cut effectively removes stars with a velocity close to the local standard of rest:
the thin disc, the majority of the thick disc, and some halo stars with disc-like dynamics.
We classify all of these removed stars as \Disc.
Likewise, our kinematically defined \Halo{} sample contains a {\GroupFont{Hot Thick Disc}} component 
\citep{bonaca17GaiaRevealsMetalrich, koppelman18OneLargeBlob, helmi18MergerThatLed}.
This selection results in a \Halo{} sample of $67,989$ stars and $17,115,924$ \Disc{} stars.
Approximately 1 in 250 stars is a \Halo{} star.

To avoid being computationally overwhelmed by millions of \Disc{} stars,
the 6D \Disc{} stars are randomly downsampled by a factor of 500,
reducing $\sim17$ million to a sample of $34,231$.
Due to the compactness of the \Disc{} in IoM space, its region of phase space is still very well populated.
This downsampling is compensated for later in the method with a weighting factor of $500$.

In this work, the effects of measurement uncertainties are neglected,
instead trusting that quality cuts and a local distance cut are sufficient for an accurate sample.
Any observational uncertainties would be statistically expected
to reduce the clumping of substructures instead of inducing or enhancing them.
The uncertainties have already broadened the halo substructure seen within the 6D RVS sample.

\subsection{Stellar halo groups}%
\label{SubSec:6D_Groups}

The \Halo{} sample is decomposed into dynamical groups identified in \citetalias{dodd23GaiaDR3View}.
This work built on the previous studies of 
\citet{lovdal22SubstructureStellarHalo} and \citet{ruiz-lara22SubstructureStellarHalo},
applying a clustering algorithm to identify statistically significant overdensities in the space of energy and angular momentum components
to find clusters of stars in their \Halo{} sample.
These clusters were then joined to form groups using a single linkage algorithm and a distance cut based on the dendrogram derived by the algorithm.
Finally, stars located within a Mahalanobis distance of $2.2$ from each group's centre 
(which approximately corresponds to a contour containing 80\% of the original members)
have been added to the groups to include additional members.
These groups are thus likely to be conservative, although not necessarily pure \citep[such as \Thamnos; see][]{dodd22GaiaDR3View}.

We use the original groups of \citetalias{dodd23GaiaDR3View} with a few minor modifications.
Several ``loose'' clusters\footnote{Clusters 34, 48, and 77 in \citetalias{dodd23GaiaDR3View}}
that fall within the Mahalanobis distance of \GES{}
but fall slightly over the dendrogram cut used in \citetalias{dodd23GaiaDR3View} to define the group
are joined to \GES.
We choose not to include the \GroupFont{LRL3} group, as it is a low-energy structure of mixed populations with an unclear origin.
Instead, its' stars are classified as \Ungrouped.
We believe that \LRLsixtyfour{} is the same as the structure Antaeus identified in~\cite{oria22AntaeusRetrogradeGroup},
and we chose to use the original name designated in~\cite{ruiz-lara22SubstructureStellarHalo}.
The membership of \EDtwo{} is updated to members identified in \citet{balbinot23ED2ColdNot},
increasing from 32 to 34 members within our RVS sample.
Similarly, the membership of \Typhon{} is updated to members identified in \citet{tenachi22TyphonPolarStream}.
The final groups used are \GES, {\GroupFont{Hot Thick Disc}} (\HTD), \Thamnos,
\GroupFont{Helmi Streams} (\HS), \Sequoia, \LRLsixtyfour, \EDtwo, \EDthree, \EDfour, \EDfive, \EDsix,
and \Typhon.
Their exact populations can be found in the legend of Fig.~\ref{fig:Data_Dodd23}.
This classification leaves approximately 57,000 \Ungrouped{} stars, which are treated as another group.

\subsection{Crossmatched sample}\label{Subsec:Crossmatch}
As an additional test dataset, we create a sample of 6D stars using
the 5D \Gaia{} dataset, also within $2.5\kpc$, crossmatched with a collection of other surveys providing \vlos{}.
This sample will allow us to test our methodology with stars entirely outside the RVS catalogue
and a comparison with the results of \citetalias{naik24MissingRadialVelocities}.
We include crossmatches of 
\textsc{APOGEE} DR17 \citep{majewski17ApachePointObservatory,abdurrouf22SeventeenthDataRelease},
\textsc{LAMOST} DR7 low and medium resolution \citep{cui12LargeSkyArea,zhao12LAMOSTSpectralSurvey},
\textsc{GALAH} DR3 \citep{buder21GALAHSurveyThird},
and \textsc{RAVE} DR6 \citep{steinmetz06RadialVelocityExperiment,kunder17RadialVelocityExperiment}.

We process the data the same as our RVS sample,
applying the same $2.5\,\mathrm{kpc}$ distance cut and similar quality cuts.
We still require an error on the crossmatched error \vlos{} less than $20\kms$,
but do not require $\left(G_{\mathrm{RVS}}-G\right)<-3$
or a signal-to-noise spectra cut.
We remove stars that are already in our 6D RVS dataset and those that are unbound.

These selections give $1,455,638$ total stars;
4,618 from \textsc{RAVE},
35824 from \textsc{APOGEE},
22,390 from \textsc{LAMOST} MRS,
1,381,077 from \textsc{LAMOST} LRS,
and 11,729 from \textsc{GALAH};
this sample is dominated by LAMOST LRS.
Using a $210\kms$ Toomre velocity cut,
this sample is classified into $1,425,886$ \Disc{} stars and $29,737$ \Halo{} stars.
\Halo{} stellar groups are defined using the \citetalias{dodd23GaiaDR3View} sample
and a Mahalanobis distance of $2.2$ (see \citetalias{dodd23GaiaDR3View} for further details.).

\section{Density in action-angle space}%
\label{Sec:ActionSpace}

This work uses axisymmetric action-angles to describe how stars are dynamically distributed.
In action-angle space, $\left(\J,\boldsymbol{\theta}\right)$,
the actions $\boldsymbol{J} = \left(J_R,J_z,J_{\phi}\right)$ describe the orbit,
and the angles $\boldsymbol{\theta}$ describe where the star is along the orbit.
The radial action $J_{R}$ and vertical action $J_{z}$ quantify the radial and vertical oscillations of orbit, respectively,
whilst the tangential action $J_{\phi}$ is equivalent to the $z$ component of angular momentum, $J_{\phi}=L_{z}$.
The corresponding angles increase linearly with time modulo $2\pi$
as the orbit completes an oscillation in the respective dimension~\cite[see further details in][]{binney08GalacticDynamicsSecond}.

In a phase-mixed galaxy, stars are spread out throughout their orbits,
uniformly distributed throughout the ${\left(2\pi\right)}^3$ volume of angle-space.
The density in action-angle space is then directly proportional to the density in action space,
divided by the volume of angle-space within which it is distributed: 
\begin{equation}\label{Eq:PhaseMixed}
  \FJt={\left(\frac{1}{2\pi}\right)}^{3}\FJ.
\end{equation}

It is this assumption and effective reduction in dimensions that simplifies our methodology
and improves the relationship between the observable 5D coordinates and the likely \vlos{}.
In the following Sec.~\ref{Sec:Method}, 
the density \FJt{} is related to the likelihood of an individual star's orbits.

\subsection{Selection function correction}%
\label{SubSec:SFCorrection}

Our sample of 6D stars suffers from several selection effects that distort the observed distributions of dynamical properties 
from their true distributions.
For a star to be in our sample, it must be: first, in the \Gaia{} RVS catalogue;
second, within $2.5\,\mathrm{kpc}$ of the Sun; and third, satisfy our quality cuts.
Only orbits that enter our solar neighbourhood of $2.5\,\mathrm{kpc}$ can be observed;
a circular orbit at $5\kpc$ or $11\kpc$ will never be seen.
For this reason,
our estimation of the orbital space in regions beyond our 6D sample's $2.5\kpc$ distance limit is inherently incomplete,
and our methodology cannot be correctly applied to stars outside our local region.

\begin{equation}
  \FJs = \FJ \,\,  \forall\,\, \J\,\, \text{that cross the Solar neighbourhood}.
\end{equation}

Even if the Galaxy is perfectly phase-mixed, our sample of observed stars will not be uniformly distributed in angle-space.
The solar neighbourhood selection in physical space corresponds to restricted regions of the angle-space that depend on the orbit's nature.
These regions are a fraction of the total ${\left(2\pi\right)}^3$ volume and
are equivalent to the fraction of time a star on an orbit is observable.
Furthermore, due to the effective magnitude cut of the \Gaia{} RVS selection function,
our sample is biased towards observing more stars closer to us.
At greater distances, we see only the intrinsically bright members of the stellar population,
whereas at closer distances, we see a larger fraction of the stellar population.

Combining our selection effects, the action distribution that we observe, \FobsJ,
is changed from the true action distribution \FJs{} of the stars that cross the solar neighbourhood.
We denote the expected observable fraction of a typical Halo stellar population on an orbit 
\J{} as  $\tau\left(\boldsymbol{J}\right)$.
The observed action distribution \FobsJ{} is then:

\begin{equation}
\FobsJ \approx \tau\left(\boldsymbol{J}\right)\ \FJs.
\end{equation}

For example, if there are 1000 stars on an orbit that crosses the solar neighbourhood, 
and the expected observable fraction is $\tau\left(\boldsymbol{J}\right)=1/100$,
we expect to observe only 10 stars on average.
The $1000$ stars are approximately evenly spread in angle about the orbit,
the majority in different parts of the Galaxy, and cannot be observed.
From an observer's perspective, we see the inverse;
we observe 10 stars and statistically expect there to be $1000$ stars upon that orbit.
We must estimate \tJ, then \FJs{} can be approximately recovered from \FobsJ{} as:

\begin{equation}\label{Eq:FJCorrected}
  \begin{aligned}
\FJs &\approx \frac{1}{\tau\left(\boldsymbol{J}\right)}\FobsJ = w\left(\boldsymbol{J}\right)\FobsJ =\FobsJcorrect\\
 &\text{for } \tau(\J) \neq 0.
  \end{aligned}
\end{equation}
where the inverse fraction of time is written as the correction weight $w\left(\boldsymbol{J}\right)$,
and the corrected observed action distribution \FobsJcorrect.
We emphasise that this approximation only holds for orbits that cross the solar neighbourhood and are possible to observe ($\tJ\neq0$).

These weights can be considered a correction to unbias \FobsJ{} from our observational,
solar-centric view of our Galaxy.
Combined with our phase-mixed assumption, the underlying phase-space density \FJt{} can be estimated\footnote{This is equivalent to assuming that the density \FJt{} can be approximated as the number of observed stars on an orbit \J{}
divided by the observable volume instead of the full $\left(2\pi\right)^3$}
using \FobsJcorrect,
for orbits \J{} that cross the solar neighbourhood.
We briefly overview of how \tJ{} is estimated below;
for further technical details of the calculation, see Appendix~\ref{Appendix:SF}.

First, for an orbit \J, we sample many physical points along its trajectory.
Next, we define a typical halo stellar population with an isochrone,
using a Chabrier \citep{chabrier03GalacticStellarSubstellar} Initial Mass Function
and an alpha-enhanced BaSTI isochrone \citep{pietrinferni21UpdatedBaSTIStellar},
with a metallicity of $\left[\mathrm{Fe}/\mathrm{H}\right]=-1.1$ and an age of $11$ Gyrs.
Then, the probability of observing a star of a given magnitude and colour from the stellar population at the sampled positions is evaluated.
This probability is estimated 
using the dust maps of \citet{lallement22UpdatedGaia2MASS3D}, photometric corrections of \citet{gaiacollaboration18GaiaDataRelease_HR},
and the RVS selection function of \citet{castro-ginard23EstimatingSelectionFunction}.
Observational uncertainties and the corresponding quality cuts are estimated using the \textsc{PyGaia}\footnote{https://github.com/agabrown/PyGaia} software package.
Finally, by averaging over the defined halo stellar population and over the points around the orbit,
the expected observable fraction \tJ{} of stars upon the orbit \J{} is found.

\subsection{Density estimation}\label{SubSec:Method_Density}

Using the phase-mixed assumption and the selection function correction,
finding the underlying phase-space density \FJt{} reduces to evaluating the corrected density \FobsJcorrect.
We model this as a collection of points and weights 
$\left[\boldsymbol{J}_i,\boldsymbol{w}_i\right]$,
from which we want to estimate the density at arbitrary \J.%
\footnote{ It is at this point that the downsampling of the \Disc{} group is compensated
by scaling the weights of \Disc{} stars by a factor of 500.
This downsampling gives a smooth, well-sampled density with a computationally practical number of points.}
This is numerically challenging, as the density varies by orders of magnitude throughout the action space.

We choose to use a Modified Breiman Estimator (MBE), as described in~\cite{ferdosi11ComparisonDensityEstimation}
and used in other works to estimate action space density \citep{sanderson15ActionspaceClusteringTidal}.
The MBE uses an Epanechnikov kernel with an adaptive smoothing length for each point to ensure a smooth density.
This method is further modified to include the selection function correction weights
and a scaled action space to ensure a smooth density for all groups.
Furthermore, to help with the numerical calculation, we first transform from action space $\left(J_R,J_z,J_{\phi}\right)$ 
to the space of $\left(\sqrt{J_R},\sqrt{J_z},J_{\phi}\right)$. 
The method is described in detail in the appendix (see App.~\ref{Appendix:Density}).

This methodology of weighted corrections and smoothing does have limitations.
We cannot estimate the density in regions of action space where we do not have any stars.
Lower binding energy orbits (equivalently larger action) have longer orbital periods and spend less fraction of their time in the solar 
vicinity, and so get larger correction weights.
This region is also where there are fewer stars due to the nature of the action distribution of the halo.
As a result, in these regions, the observed distribution is very poorly sampled and hence more stochastic,
relying on the few observed stars to receive a high correction weight, which is then given a very large smoothing.
These areas do not give very reliable density estimates for the halo.

\section{Method}\label{Sec:Method}

\begin{figure*}[htp!]
\label{fig:J_path}
 \includegraphics[width=\textwidth]{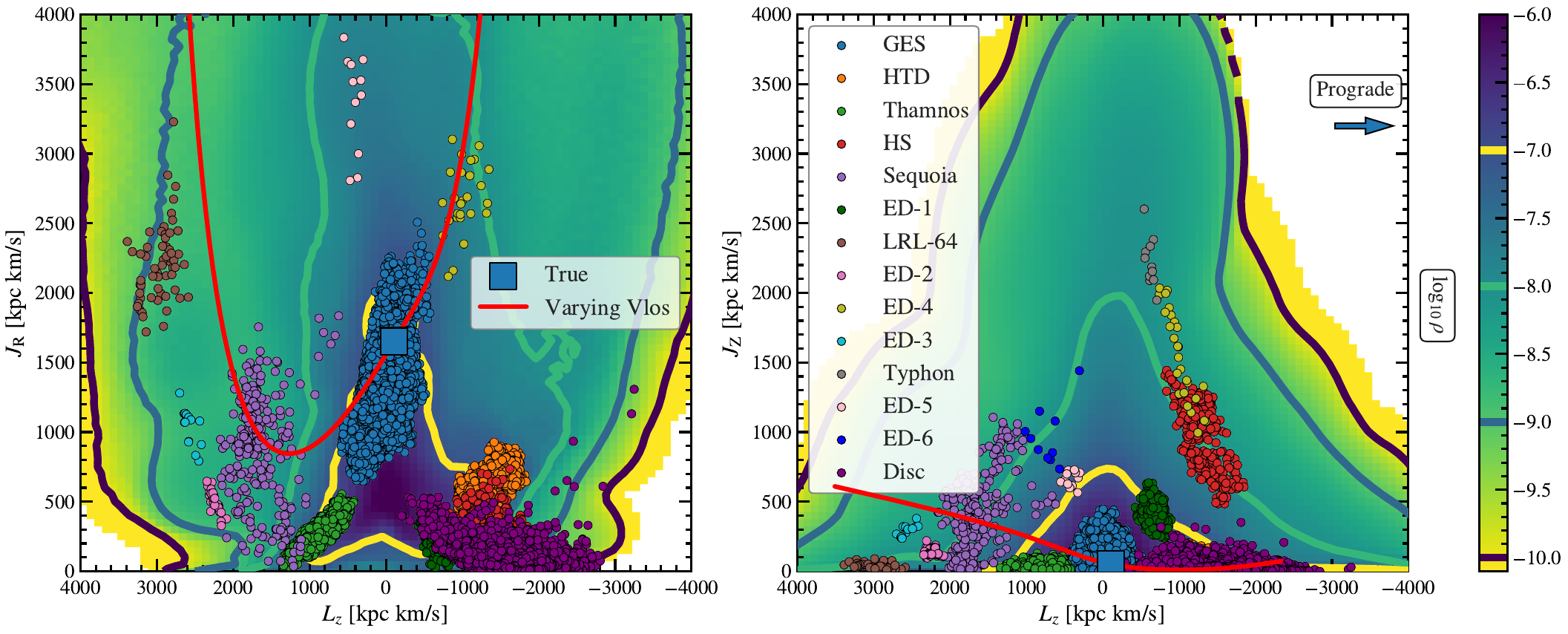}
 \vspace{-12pt}
 \caption{Our sample of 6D \Gaia{} DR3 RVS stars in action space,
  classified into different groups of {\protect{\citetalias{dodd23GaiaDR3View}}} (indicated by colour).
  The \Ungrouped{} stars are represented by smoothed density,
  as indicated by the colours and contours in the colour bar.
  This density, and those of the individual groups,
  are corrected for selection effects (see text).
  Across the space is an example path (red line) of a 5D star through action space found
  by varying an assumed \vlos, with values indicated in the legend.
  The corresponding PDF can be seen in the following Fig.~\ref{fig:ex_pdf}.
  This example star is a \GES{} member from the 6D \Gaia{} RVS sample,
  with the true \vlos{} and the corresponding point in action space marked by a square.
 \vspace{-10pt}
 }
\end{figure*}

\begin{figure}[h!]
 \includegraphics[width=\columnwidth]{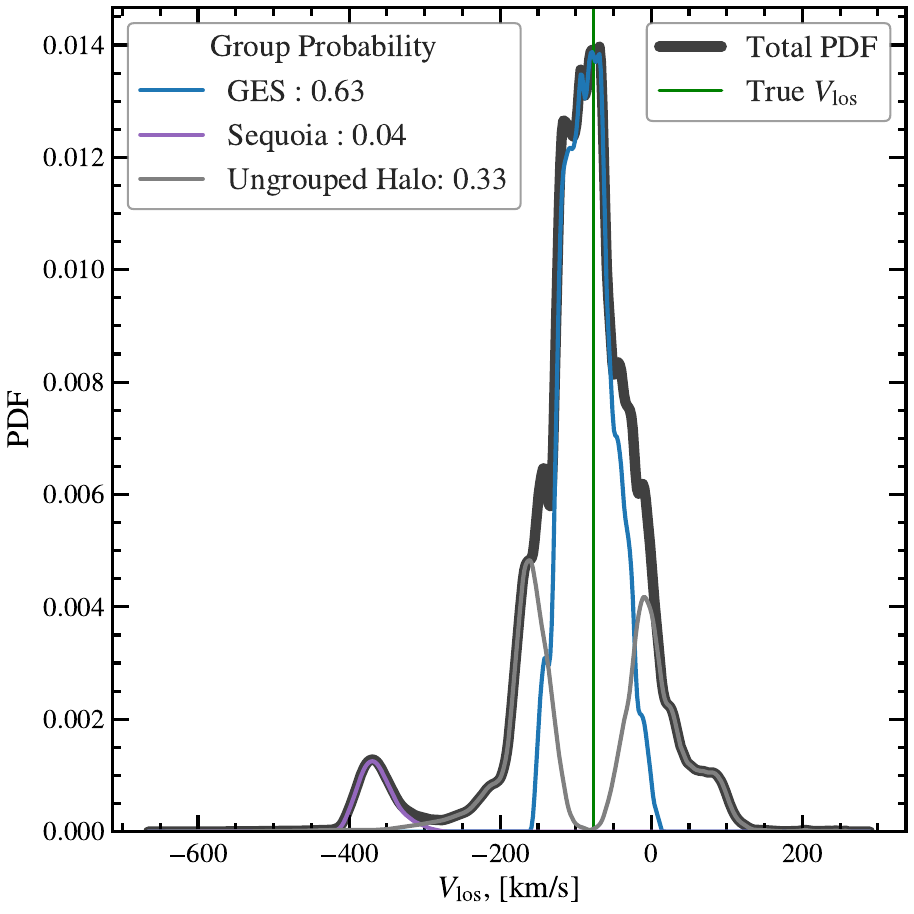}
 \vspace{-12pt}
 \caption{PDFs of \vlos{},
   calculated from a 5D star's path in action space (see Fig.~$2$). 
  The thick black line total PDF shows the total density.
  This total can be decomposed into contributions from different groups, denoted by different colours.
  The probability that the star belongs to these different groups can be found by integrating the area under the respective lines,
  shown in the legend.
  As a star from the 6D sample, the true \vlos{} is marked in green, and it is known that it is a \GES{} member.
  Our method finds a $63\%$ probability of belonging to \GES{}
  but could also belong to \Ungrouped{} or \Sequoia.
  }%
 \vspace{-20pt}
\label{fig:ex_pdf}
\end{figure}

A star is observed with 5D coordinates denoted $\boldsymbol{u}$: parallax, position on sky, and proper motions.
Given this information, we want to infer the probability of the star having a particular \vlos,
the probability density function (PDF) denoted $p\left(v_{\mathrm{los}}|\boldsymbol{u}\right)$.
The PDF can be expressed as:
\begin{equation}
 p\left(v_{\mathrm{los}}|\boldsymbol{u}\right) =
 \frac{p\left(v_{\mathrm{los}},\boldsymbol{u}\right)}{p\left(\boldsymbol{u}\right)}
 \propto p\left(v_{\mathrm{los}},\boldsymbol{u}\right).
\end{equation}
As only the \vlos{} is varied, the normalisation factor of 
${p\left(\boldsymbol{u}\right)}$ can be dropped,
provided that, at the end of the calculation, the PDF is renormalised so that it integrates to unity over the space of possible \vlos.

The transformation from observed space $\left(v_{\mathrm{los}},\boldsymbol{u}\right)$
to 6D Galactocentric cartesian space $\left(\boldsymbol{x},\boldsymbol{v}\right)$ uses fixed solar parameters.
The Jacobian factor is independent of the \vlos{},
as the \vlos{} can be considered as a velocity component of a rotated, shifted, and boosted 
cartesian frame.
Again, this Jacobian factor can be treated as a normalisation factor and dropped:
\begin{equation}
 p\left(v_{\mathrm{los}},\boldsymbol{u}\right) \propto p\left(\boldsymbol{x},\boldsymbol{v}\right).
\end{equation}

The transformation between \Cart{} and \Act{} is canonical, and so:
\begin{equation}
 p\left(\boldsymbol{x},\boldsymbol{v}\right) = \FJt.
\end{equation}

\noindent
Here, \FJt{} is the distribution of \textit{all} stars in action-angle space.
Collecting everything together, it follows that:
\begin{equation}
 p\left(v_{\mathrm{los}}|\boldsymbol{u}\right) \propto \FJt.
\end{equation}

The normalised \vlos{} PDF is then:
\begin{equation}
 p\left(v_{\mathrm{los}}|\boldsymbol{u}\right)
= \frac{1}{\lambda\left(\boldsymbol{u}\right)}
\FJt|_{\left(\vlos,\boldsymbol{u}\right)},
\end{equation}
where the total normalisation factor is:
\begin{equation}
\lambda\left(\boldsymbol{u}\right) = \int \FJt|_{\left(\vlos,\boldsymbol{u}\right)} \operatorname{d}\vlos,
\end{equation}
depending on the path through the action-angle space set by the 5D coordinates.
This path is evaluated at \vlos{} between the positive and negative values
that correspond to the \vlos{} that brings the star's total speed to escape velocity.

Using results from Sec.~\ref{Sec:ActionSpace},
by assuming that Galaxy is phase-mixed, the density in \Act{} space can be assumed to be directly proportional to the density in action space \FJ{}
(Eq.~\ref{Eq:PhaseMixed}).
For \J{} that cross into our solar neighbourhood, this can be approximated by 
our observed sample of stars, corrected for selection effects, \FobsJcorrect{} (see Sec.~\ref{SubSec:SFCorrection}, Eq.~\ref{Eq:FJCorrected}).
Then:
\begin{equation}
 p\left(v_{\mathrm{los}}|\boldsymbol{u}\right) \propto\FobsJcorrect.
\end{equation}
Note that this only holds for orbits \J{} that cross the solar neighbourhood due to the selection of our observed 6D sample.
If the 5D stars position lies within our $2.5\kpc$ selection solar selection, this is assuredly the case.
The following  descriptions refer to the path and density as within action space,
but this is a generalisation of assuming a phase-mixed action-angle space.

The shape of a star's \vlos{} path through the action space depends on the observed coordinates $\boldsymbol{u}$.
The lowest point in action space (nearest to the origin) approximately corresponds to the minimum possible energy of the orbit.
As the potential energy depends on the fixed position,
the minimum total energy is where the \vlos{} minimises the kinetic energy.
This value is not necessarily at $\vlos=0$
due to the Sun's motion in the coordinate transformation from heliocentric coordinates.
As the magnitude of \vlos{} increases to become the dominant component,
the total velocity and kinetic energy also rise, the actions increasing until the orbit becomes unbound, with $E>0$.
Here, the actions are undefined, and the density and corresponding likelihood are zero.

An example of applying our method to a star from our 6D sample
can be seen in Fig.~2, 
depicting the \vlos{} path through action space,
and the corresponding (total) PDF in Fig.~\ref{fig:ex_pdf}.
The path that this particular star takes crosses the region of action space associated with \GES,
but also traverses the region of \Sequoia{} and \Ungrouped.
In this case, the region of \GES{} is the densest, and so the star is most likely to have a \vlos{}
corresponding to an orbit consistent with \GES.
For this chosen example, this is close to the true \vlos{} of the star.

\subsection{Group membership}\label{SubSec:Method_Group}
The total density can be decomposed into contributions from each of the stellar Halo substructures' distributions:

\begin{equation}
\FJt = \sum_{g\in\mathrm{Groups}} F_{g}\left(\boldsymbol{J},\boldsymbol{\theta}\right).
\end{equation}

\noindent
The individual group's PDFs are then normalised using the integral of the total PDF:
\begin{equation}
 p_{g}\left(\vlos|\boldsymbol{u}\right)=
 \frac{1}{ \lambda\left(\boldsymbol{u}\right)}
F_{g}\left(\boldsymbol{J},\boldsymbol{\theta}\right)|_{\vlos,\boldsymbol{u}},
\end{equation}
so that the total PDF integrates to one.
The integral under the group's individual PDFs gives the membership probability
of the star with 5D coordinates $\boldsymbol{u}$ belonging to each group $g$:
\begin{equation}
p_{g}\left(\boldsymbol{u}\right) = \int p_{g}\left(\vlos|\boldsymbol{u}\right) d\vlos.
\end{equation}
By construction, these probabilities sum to one; the stars must belong to one of the groups 
(including the \Ungrouped).
The total \Halo's PDF and probabilities are simply the sum of all groups except the \Disc.

In action space, there is a small region of overlapping \Halo{} and \Disc{}
where a given orbit's velocities within the solar neighbourhood
can variously fall just above or below the $210\kms$ Toomre velocity cut.
To ensure consistency with our initial kinematic definitions of \Halo{} and \Disc{},
a Toomre velocity cut is applied to the tested stars.
For a given \vlos, if the star has a $V_{\mathrm{Toomre}}>210\,\kms$, then any density from the \Disc{} is 
moved to the \HTD{}, or \Ungrouped{} if there is not already contribution from the \HTD{},
with the \Disc{} density set to zero.
Similarly, a star with $V_{\mathrm{Toomre}}<210\,\kms$ has any contribution from the \Halo{} components moved to the \Disc.
This criteria only affects a minimal selection of orbits
but prevents the inference of a \Disc{} or \Halo{} membership that does not satisfy our kinematic definition.

The individual group PDFs and probabilities can be seen for the example 5D star in Fig.~\ref{fig:ex_pdf}.
As stated before, \GES{} is the dominant contribution to the density across the path for this star, with a probability of $63\%$.
However, there are also probabilities that the star belongs to \Sequoia{} or the \Ungrouped.
There is zero probability of this star belonging to the \Disc{} as there is no possible disc-like orbit for any \vlos{}
because the path in action space is far from the region associated with the disc.

\subsection{Phase-clumped and partially phase-mixed substructures}\label{SubSec:phase-mixed}

\begin{figure}
 \includegraphics[width=\columnwidth]{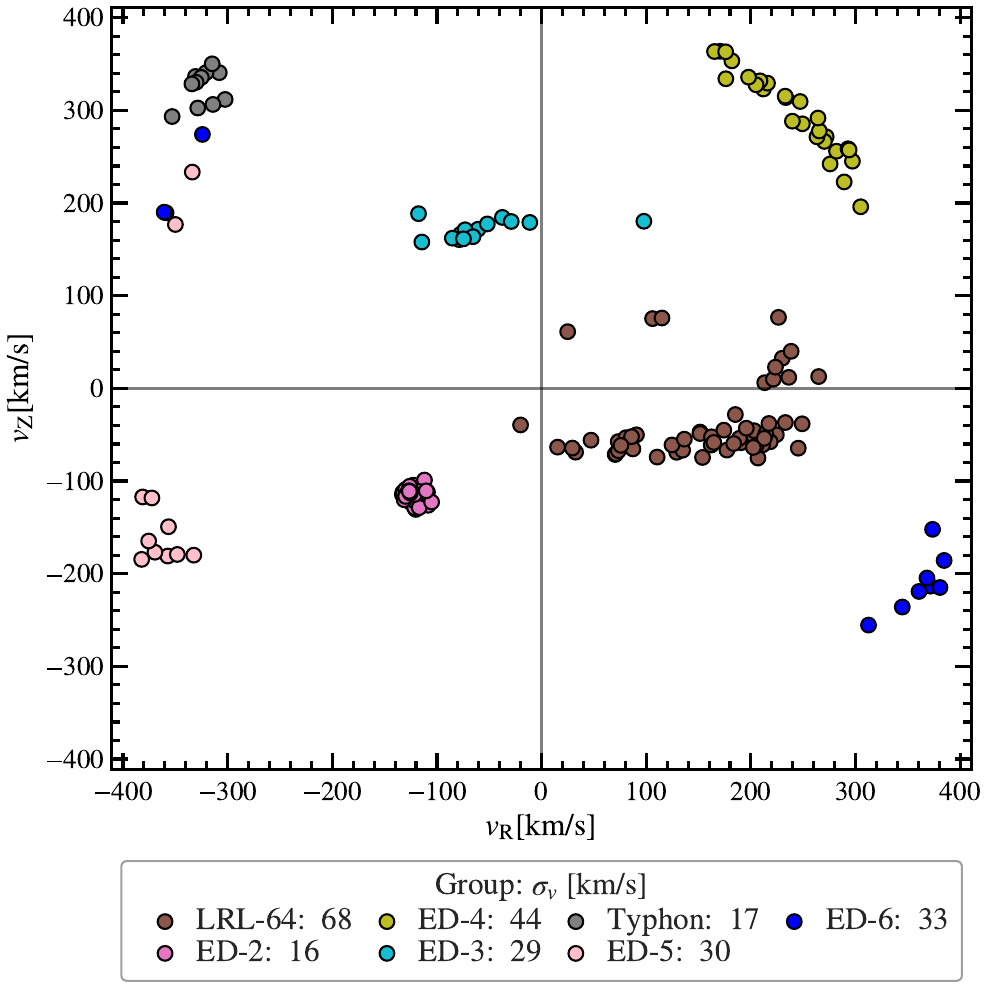}
 \vspace{-12pt}
 \caption{ Stars of un-phase-mixed groups in radial and vertical velocity space (in Galactic cylindrical coordinates).
 If the groups were phase-mixed, there would be a roughly equal number of stars from each group in every quadrant.
 Whilst not every group is in a single quadrant, none are consistent with being evenly distributed.
 }%
\label{fig:VelPhaseMix}
\end{figure}

Our methodology can be improved by recognising that not all groups are phase-mixed within the observable volume.
If a structure is truly phase-mixed, then the number of positive and negative 
values of its stars' radial and vertical velocities $(v_R,v_Z)$ are expected to be approximately evenly distributed
(where $R$ is the radius in the plane).
There should be approximately the same number of stars traveling up through the $z=0$ plane as traveling down,
and as many stars going towards and away from the Galactic centre.

In $(v_R,v_Z)$ space, this would be roughly an equal number of stars in each positive-negative quadrant for each group.
Not all of the \Halo{} substructures are phase-mixed;
the clear exceptions can be seen in velocity space in Fig.~\ref{fig:VelPhaseMix}.
The following groups are referred to as phase-clumped:
\EDtwo{} (with a particularly tight velocity spread), \EDthree, \EDfour, \EDfive, \EDsix, \LRLsixtyfour, and \Typhon.
These groups also show tight grouping in the tangential component of velocity (by construction).

The stars in these groups are phase-clumped together in a smaller volume within the larger observable volume.
As a result, these structures have higher density in some parts of the angle-space
but smaller, or zero, in the rest.
For these groups, $F_{g}\left(\boldsymbol{J},\boldsymbol{\theta}\right)$ is not well estimated by assuming 
that the stars distributed as $F_{g}\left(\boldsymbol{J}\right)$ are spread uniformly in angle-space.
Instead, we may assume these groups are uniformly distributed in a restricted volume of angle-space.

The restricted volume is defined by making an additional velocity selection,
requiring that the velocity of the evaluated point be within
a characteristic velocity length scale of known group members in velocity, $\left(v_R,v_Z,v_{\phi}\right)$.

\begin{equation}
 \eta_{g}\left(\boldsymbol{v}\right) = \begin{cases}
1, &\boldsymbol{v}\text{ is within }\sigma_{v}\text{ of a known 6D group member}\\
0, &\text{else.}
\end{cases}
\end{equation}

\noindent
The velocity length scale $\sigma_{v}$ is estimated for each group by considering the largest range $\frac{1}{2}\left(84\% - 16\%\right)$
of the distribution of different absolute velocity components,
and can be seen in the \RefChange{legend} of Fig.~\ref{fig:VelPhaseMix}.
These length scales are deliberately chosen to be generous
so as not to be overly restrictive while effectively describing the space.
These velocity criteria are applied as an additional selection function when calculating the correction weight for stars in the phase-clumped groups,
effectively reducing the volume of angle-space that the star can be observed in and increasing the correction weight.

When evaluating $F_{g}\left(\boldsymbol{J},\boldsymbol{\theta}\right)$, the position in velocity space is additionally checked:

\begin{equation}
 F_{g}\left(\boldsymbol{J},\boldsymbol{\theta}\right)|_{\vlos,\boldsymbol{u}} \approx \frac{1}{{\left(2\pi\right)}^{3}}
\eta_{g}\left(v_R,v_Z,v_{\phi}\right) F^{*}_{g}\left(\boldsymbol{J}\right)|_{\vlos,\boldsymbol{u}}.
\end{equation}

If the velocity satisfies the velocity cut, the group's contribution to the density is included;
otherwise, it is zero.
In later analysis, this additional phase-clumped step is found to
improve the recovery of these structures and reduce contamination from non-group members.
This process is only applied to the identified phase-clumped groups.

\section{Testing vlos predictions}\label{Sec:PDFs}
\begin{figure}
 \includegraphics[width=\columnwidth]{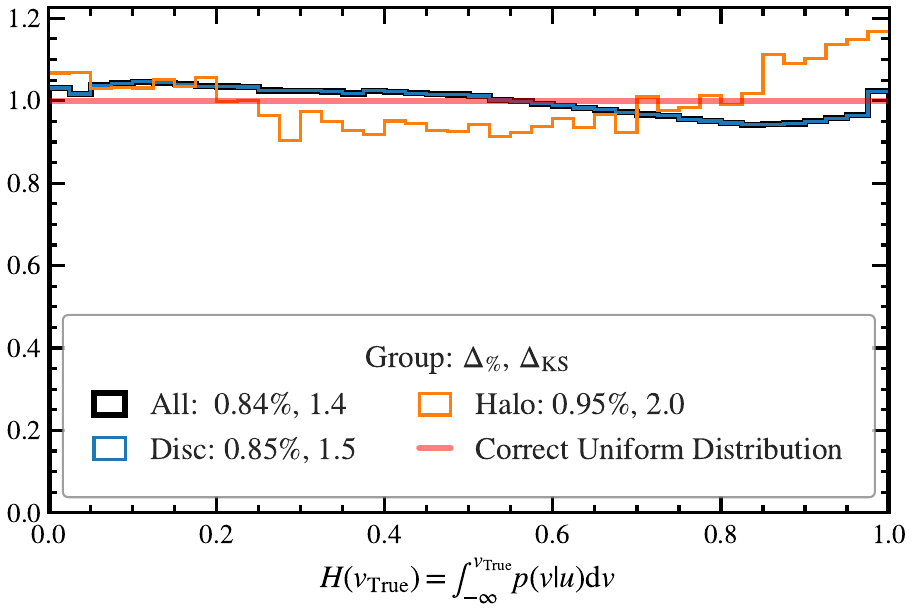}
 \vspace{-12pt}
 \caption{Distributions of CDF values $H\left(v_{\mathrm{True}}\right)$,
 calculated from the true \vlos{} of the stars and the PDFs of our method,
 for each star in our \Gaia{} RVS sample.
  If all assumptions and the PDFs are correct, these values should be distributed uniformly.
  We plot the total distribution of all stars
  and the distributions of the \Disc{} and \Halo{} stars (with CDF values corresponding to their respective PDFs).
  The total statistical error $\Delta_{\%}$ of our PDFs can be estimated by integrating the difference between the found 
  and true uniform distribution.
  \vspace{-10pt}
 }%
\label{fig:6D_VlosCDF}
\end{figure}

We now test our methodology on both our 6D samples.
Every star's 5D information is used to predict a \vlos{} PDF and group membership probabilities.
These predictions can then be evaluated using the true \vlos{} and the group membership of \citetalias{dodd23GaiaDR3View}.
For the RVS sample, the stars define the action-angle density used within the method.
Therefore, when testing a star from this sample, the density contribution from its own true point is removed
when evaluating the density along its path in action space.
This measure ensures that the test is not biased favourably, with an advantage that cannot be replicated when applying to true 5D stars.
In practice, this makes a negligible difference due to the density smoothing.

In the following tests and discussions,
the samples are frequently decomposed into \Halo{} and \Disc{} stars for separate analysis.
We find that predicting the \vlos{} for these kinematic groups is very different,
as it is generally far easier to constrain a \Disc{} star to a \Disc{} orbit
than a \Halo{} star to a specific \Halo{} orbit (as shown in the following section).
Without considering these groups separately, the total sample and results are dominated by \Disc{} stars,
masking any meaningful study of the \Halo.

The statistical accuracy of the inferred \vlos{} PDFs can be quantified by considering the cumulative distribution function (CDF, defined $H$)
evaluated at the true \vlos:
\begin{equation}
 H_{\mathrm{True}} = H\left(v_{\mathrm{True}}|\boldsymbol{u}\right)
 =\int_{-\infty}^{v_{\mathrm{True}}}p\left(v|\boldsymbol{u}\right)\mathrm{d}v.
\end{equation}
If the predicted PDFs are genuinely accurate and unbiased, the total distribution of $H_{\mathrm{True}}$
should be uniformly distributed\footnote{
Consider, if a single distribution $p\left(\vlos|\boldsymbol{u}\right)$ is fairly sampled many times,
it is expected that only $X\%$ of the drawn \vlos{} should be in the bottom $X\%$ of the distribution for all $X$.
This principle also holds for values individually fairly drawn from many different
$p\left(\vlos|\boldsymbol{u}\right)$ distributions.}.
Similarly, a subselection in any quantity independent of \vlos, such as distance and positions,
should be uniformly distributed.

However, 
selecting kinematically defined groups of stars includes an implicit selection on \vlos{} and is,
therefore, a biased selection of $H_{\mathrm{True}}$.
Instead, cumulative distribution values are calculated from a group's own PDF rather than the total PDF.
These CDF values are then selected and calculated with the knowledge that they are true group members
and are statistically expected to be distributed uniformly.
The total distribution can then be decomposed into separate \Disc{} and \Halo{} distributions,
ensuring that the predicted PDFs are statistically accurate for both groups of stars.

The distributions of the cumulative values for the total \Gaia{} RVS sample%
\footnote{Without downsampling the disc.}
are plotted in Fig.~\ref{fig:6D_VlosCDF},
alongside the \Disc{} and \Halo{} subsamples.
These distributions are nearly, but not quite, uniform.
The total is, as expected, dominated by the \Disc{} distribution, which shows a clear slight asymmetric trend.
We attribute this trend to disequilibrium within the \Disc{}
due to the correlation between dynamical structure and small systematic under and overestimating of \vlos{}
throughout the solar neighbourhood (see App.~\ref{App:Disc} for details).
This disequilibrium violates our assumptions of phase-mixed, axisymmetric equilibrium.
The \Disc{} PDFs are very narrow, and as a result, these minor biases in the CDF only correspond to a few \kms{} in \vlos{}
and are not detectable in other tests and analyses.

The \Halo{} distribution shows a slight excess of stars near cumulative values $0$ and $1$,
indicating that more true \vlos{} values are in the tails of the predicted PDFs than is statistically expected.
The tails of the PDFs are typically regions of high \vlos{}, corresponding to areas of high energy and action.
This excess suggests that the density in these regions is slightly underestimated,
likely due to the difficulty in reliably estimating the density in these areas (see Subsec~\ref{SubSec:Method_Density}.).

The differences between our results and the correct uniform distributions are quantified with two statistics.
First, following \citetalias{naik24MissingRadialVelocities},
the total absolute area between the found distribution $f\left(H\right)$ and the ideal uniform distribution:
\begin{equation}
 \Delta_{\%} \approx \frac{1}{2}\sum_{i}^{\mathrm{bins}}\left|f^{\mathrm{Hist}}\left(H_i\right) -1 \right|\Delta_{x},
\end{equation}
where $\Delta_{x}$ is the width of the bins.
Note that this statistic depends upon the choice of binning and Poisson noise in smaller samples,
such as the \Halo{} sample.
In addition, the KS statistic $\Delta_{\mathrm{KS}}$ describes the maximum difference between the cumulative distributions,
scaled by a factor of 100 for convenience:
\begin{equation}
 \Delta_{\mathrm{KS}} =100\times\max \left|
H^{\mathrm{Hist}}\left(H_{\mathrm{True}}\right) - H^{\mathrm{Uniform}}\left(H_{\mathrm{True}}\right) \right|.
\end{equation}
This quantity has the advantage that it can be calculated without a choice of bins and is more robust to small number statistics.

For the \Disc{} distribution,
we find $\Delta_{\%}=0.85\%$ and $\Delta_{\mathrm{KS}}=1.4$,
whilst the \Halo{} has values of $\Delta_{\%}=0.95\%$ and $\Delta_{\mathrm{KS}}=2.0$.
These statistics confirm that both populations are well recovered, 
without strong systematic biases.

\subsection{Constraining vlos}\label{Subsec:ConstrainVlos}
\begin{figure}
 \includegraphics[width=\columnwidth]{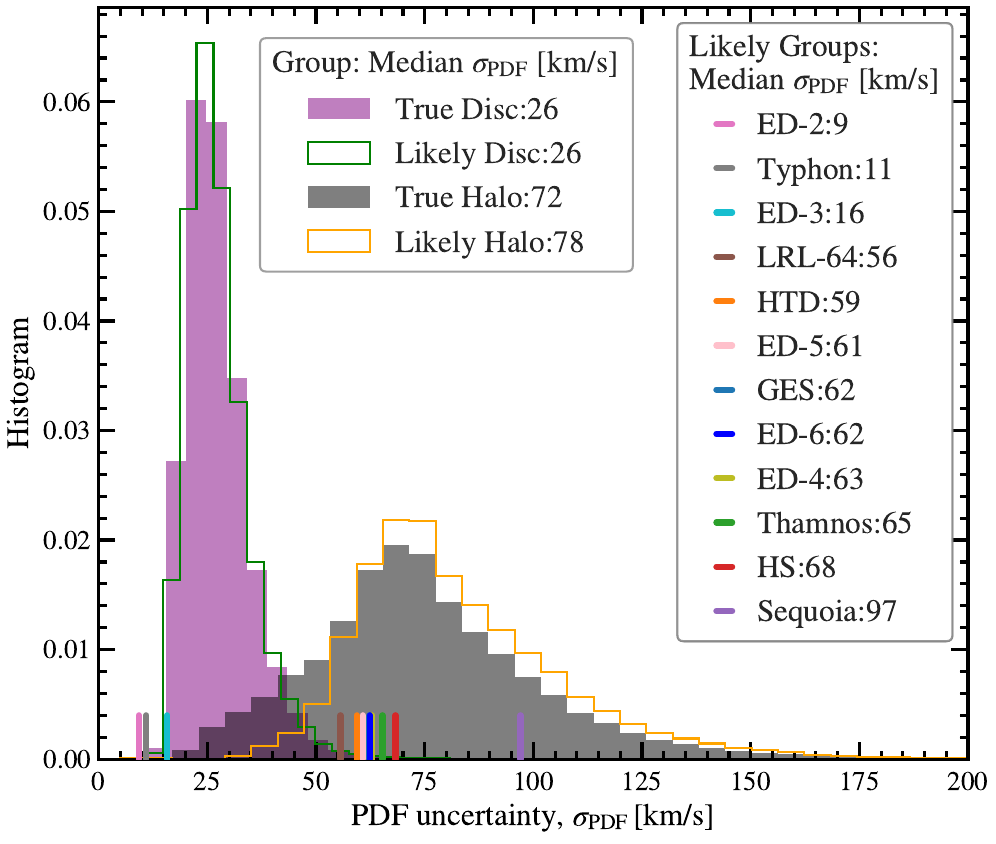}
 \vspace{-12pt}
 \caption{
 Distributions of $\sPDF$ (defined as half the $84\%-16\%$ \vlos{} values) of the PDFs predicted by our method.
 The sample is split into \Halo{} and \Disc{} stars,
 whose normalised histograms of the true decomposition are plotted with solid-filled colours.
 The likely decomposition, defined by the probability of membership greater than 80\%, 
 is plotted with lines.
 The median values for the distributions for stars belonging to different groups 
 (based on {\citetalias{dodd23GaiaDR3View}}) are shown in different colours.
 \vspace{-10pt}
 }\label{fig:6D_Sigma}
\end{figure}

Whilst the predicted PDFs can be considered statistically consistent with the truth,
that does not necessarily mean that they effectively constrain the likely \vlos{} of a given star.
The predictive power of our \vlos{} PDFs can be quantified by defining the uncertainty in our \vlos{} estimate:
\begin{equation}
 \sigma_{\mathrm{PDF}} = \frac{1}{2}\left(v_{84\%}- v_{16\%}\right),
\end{equation}
where $v_{X\%}$ corresponds to the \vlos{} that is $X\%$ of the way through the cumulative distribution of a given PDF, 
$X\%=H\left(v_{X\%}|\boldsymbol{u}\right)$.
The smaller the \sPDF{} range, the more constraining the PDF, and the smaller the uncertainty on a predicted \vlos{}.

The distribution of \sPDF{} for the RVS sample is plotted in Fig.~\ref{fig:6D_Sigma},
split into both the true \Disc{} and \Halo.
Stars belonging to the \Disc{} are well-constrained, with a median \sPDF{} of $26\kms$,
slightly smaller than the velocity dispersion within the thin disc \citep{bland-hawthorn16GalaxyContextStructural}.
This average uncertainty is the same value as the total sample (which is dominated by the disc)
and similar to the results of \citetalias{naik24MissingRadialVelocities}.
As shown in the following Sec.~\ref{Sec:Probs},
the overwhelming majority of \Disc{} stars are correctly identified.
Stars belonging to the general \Halo{} are far less constrained,
with a larger median \sPDF{} of $72\kms$.

However, stars likely to belong to substructure within the \Halo{} are typically better constrained,
as can be seen in the median \sPDF{} values of likely members (defined as membership probability greater than $80\%$).
In general, the likely members of smaller substructures have much smaller \sPDF{} values,
as they are so narrow and dense in the phase-space that only a small range of \vlos{} corresponds to a matching orbit.
The groups that are broad in action space typically have members with wider, less constraining PDFs.
The group with the largest median \sPDF{} is \Sequoia,
which is reasonably broad in action space and has a small membership for its size.
These combined factors result in a lower density in action space,
leading to less dominant, less sharply peaked, and less constraining PDFs.

The sample is additionally decomposed into likely \Halo{} and \Disc{} members (defined at a membership probability above $80\%$).
There is a noticeable deviation at lower uncertainty between the distributions of true \Halo{} stars and those that are found likely to belong to the \Halo{}.
A small fraction of \Halo{} stars are being misattributed to the \Disc{} with high probability,
being predicted by narrowly constrained PDFs corresponding to disc-like orbits.

\subsection{Predicting vlos}\label{SubSec:VlosPredict}

\begin{figure}
 \includegraphics[width=\columnwidth]{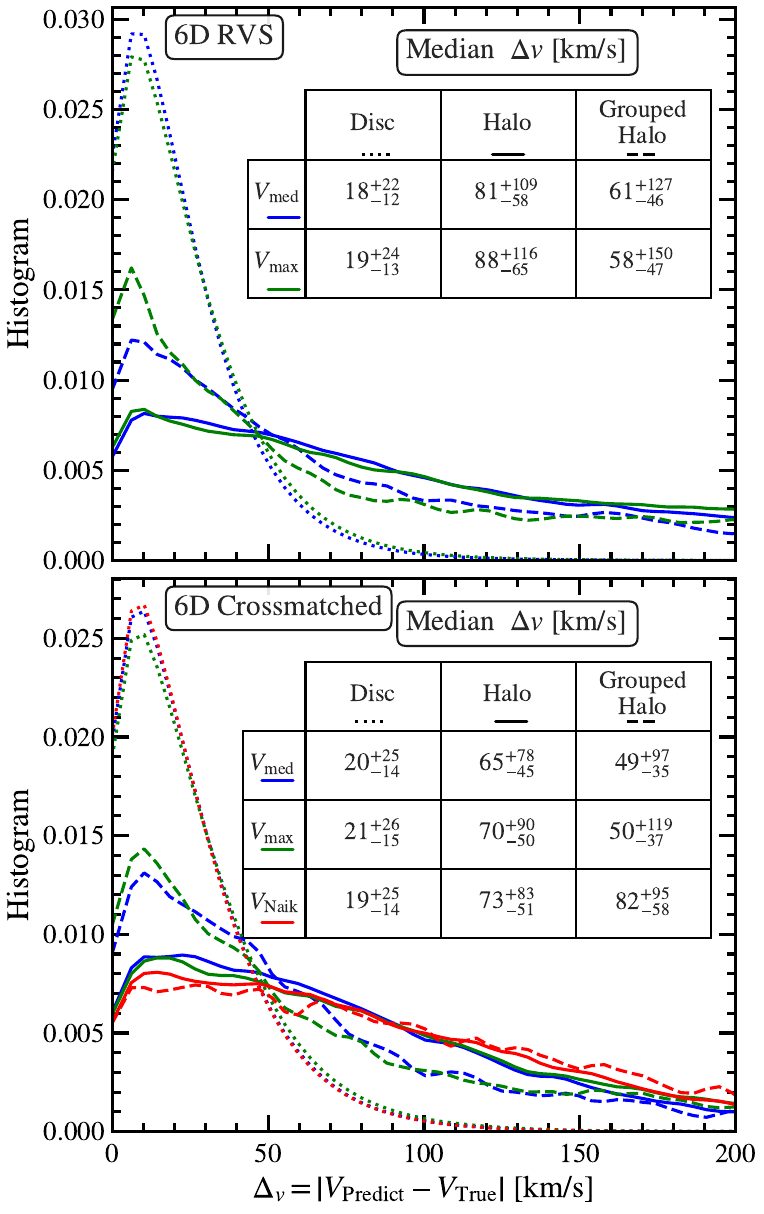}
 \vspace{-12pt}
 \caption{Distributions of the difference between \vlos{} predictions and the true \vlos{}
 across different subsamples of the \Disc, the total \Halo, and the stars belonging to substructure in the \Halo.
 The top panel sample is from the \Gaia{} RVS catalogue
 and includes \vlos{} estimates our PDF's the median and peak maximum values.
 The bottom panel shows a sample of stars from surveys crossmatched with \Gaia{}
 and includes \vlos{} estimates from \citetalias{naik24MissingRadialVelocities}.
 The tables contain the median velocity error for the different velocity predictions across the different subsamples.
 }%
 \vspace{-12pt}
\label{fig:Verror}
\end{figure}

\begin{figure}[h!]
 \includegraphics[width=\columnwidth]{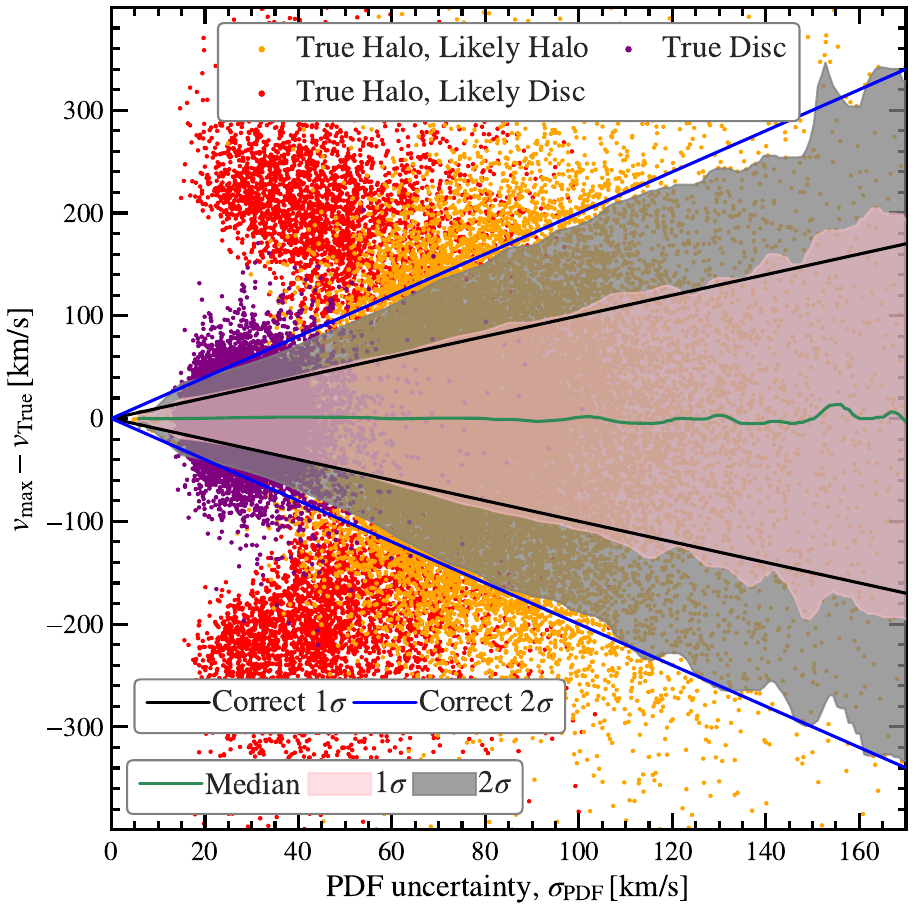}%
 \vspace{-12pt}
 \caption{
 Uncertainty in our predicted \vlos{} PDFs against  the difference between the true value $v_{\mathrm{True}}$
 and our predicted line-of-sight velocity $v_{\mathrm{max}}$.
 True \Disc{} stars are plotted in purple, downsampled by a factor of 1000, 
 the true \Halo{} stars that are determined to be likely \Halo{} stars by our method (probability greater than $80\%$) in orange,
 and true \Halo{} stars that are misdetermined to be likely disc stars in red, both downsampled by a factor of 2.
 The median line and shaded regions indicate the binned running median and $1\sigma$ and $2\sigma$ values of the total distribution,
 defined from the $16\%-84\%$ range.
 If the predictions are unbiased and the predicted PDF uncertainties accurate,
 the binned $1\sigma$ and $2\sigma$ values  regions are expected to follow the value of their own corresponding bin,
 plotted in solid black and blue lines. 
 }\label{fig:VlosPredict_VSigma}
\vspace{-10pt}
\end{figure}

Predicting a single value estimate for the \vlos{} of an individual star can be difficult,
as many of the star's PDFs are complex and multipeaked and, therefore, are not well summarised by a single value.
It is more robust to marginalise over the entire \vlos{} distribution,
as discussed in the following Sec.~\ref{SubSec:DynDist}.
If it is necessary to distil the PDF into a single \vlos{} estimate, there are two obvious choices:
the \vlos{} corresponding to the PDF's peak value (\vmax)
or the \vlos{} corresponding to the median value of the PDF ($V_{\mathrm{med}}$).
We then define the difference between our predictions and the true \vlos{} value as 
$\Delta_{v} = \left|v_{\mathrm{Predict}}-v_{\mathrm{True}}\right|$.

The distributions of the differences $\Delta_{v}$ for our predictions \vmax{} and \vmed{} are plotted in Fig.~\ref{fig:Verror},
decomposed into \Disc, total \Halo{}, and grouped \Halo{} (stars belonging to substructures).
The top panel shows the results for the RVS sample, and the bottom shows similar results for the crossmatched sample,
for which we additionally have the results of \citetalias{naik24MissingRadialVelocities}.
These results follow the distributions of Fig.~\ref{fig:6D_Sigma},
where the \Disc{} have, on average, the smallest differences $\Delta_{v}$.
The general \Halo{} stars have comparatively larger differences $\Delta_{v}$,
but likely substructure members can be found with greater accuracy.
We find that \vmax{} and $V_{\mathrm{med}}$ give similar results on average,
although the individual estimates for stars can vary.
The \vmax{} estimates perform marginally better for well-constrained PDFs with a single dominant peak,
typically found in PDFs of likely substructure members.
However, \vmax{} estimates differ more from the true \vlos{} than \vmed{} estimates on average for less well-defined PDFs.

To ensure that the differences between \vlos{} estimates and the true value are statistically consistent with the expected uncertainty \sPDF,
consider Fig.~\ref{fig:VlosPredict_VSigma}.
By binning stars in \sPDF, the running median, $1\sigma$, and $2\sigma$ of the differences between the predicted and true \vlos{}  within the bin are calculated,
defined by the percentile ranges $16$-$84\%$ and $5$-$95\%$, respectively.
If the predicted uncertainties and \vlos{} estimates are statistically accurate,
the $1\sigma{}$ of the distribution within bins should equal the \sPDF{} value of the bin itself.
We find good agreement across the entire range of \sPDF;
the uncertainties in our \vlos{} PDFs are statistically consistent with the true differences between our prediction and the true value.

By considering the subsamples of stars, indicated by colour in Fig.~\ref{fig:VlosPredict_VSigma},
the same patterns as before can be seen.
The \Disc{} stars are well recovered and constrained,
whilst the majority of the \Halo{} stars are identified with greater \sPDF{} that is statistically consistent with the true
differences between the predicted and the true \vlos{} values.
However, a small population of true \Halo{} stars is misattributed to the \Disc{} (in red).
These stars are predicted with small \vlos{} uncertainties, but, in reality,
the differences between our predicted and true \vlos{} values are large.

\subsection{Marginalising over vlos}%
\label{SubSec:DynDist}
\begin{figure*}
 \includegraphics[width=\textwidth]{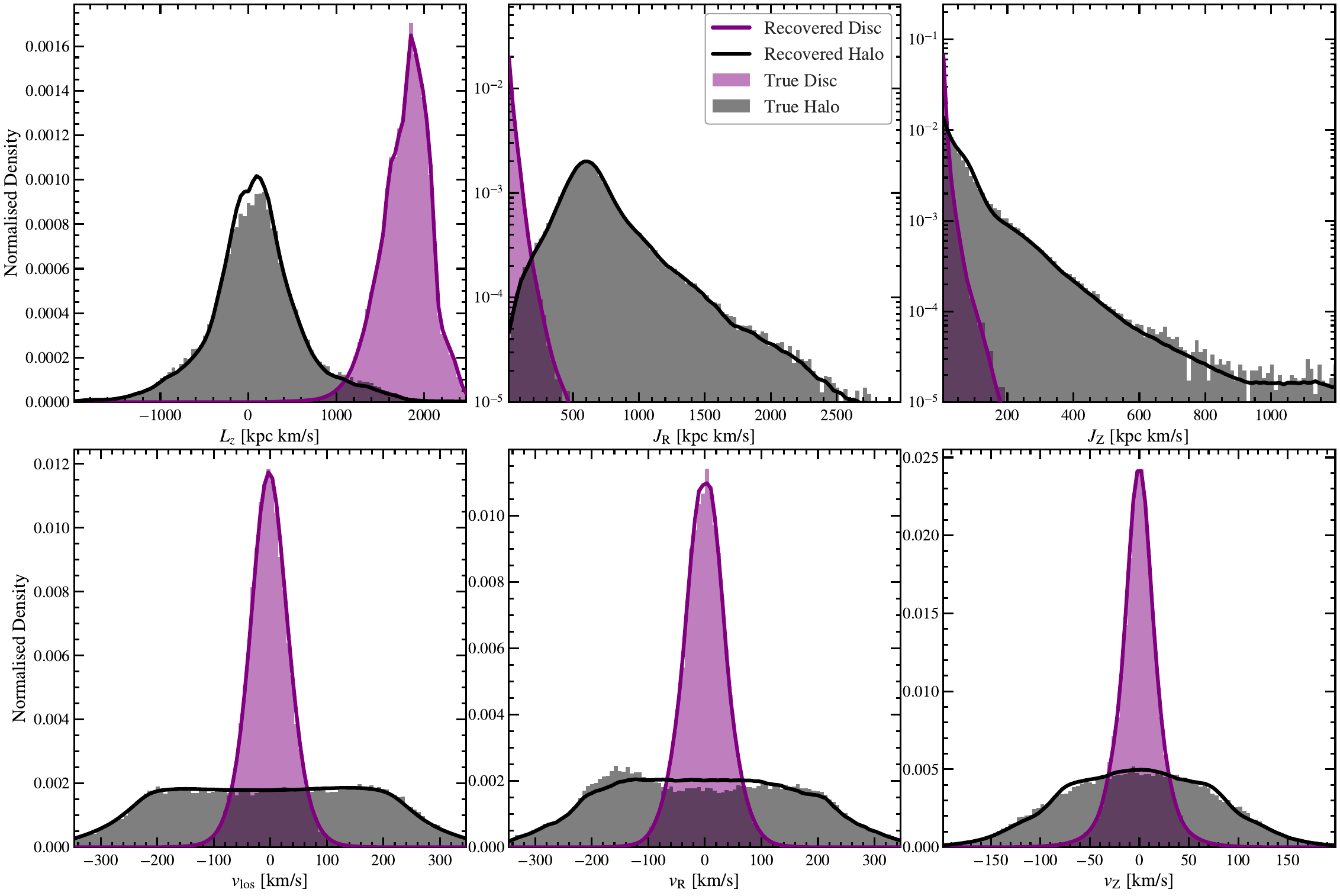}
 \vspace{-12pt}
 \caption{ Distributions of various dynamical quantities of the \Disc{} and \Halo{} samples,
 defined kinematically by $\vtoomre>210\kms$.
 The solid distributions are the original 6D distributions,
 and the lines are the recreated distributions from the 5D of the 6D sample by marginalising over the likely \vlos{} PDFs
 inferred by our method.
 The top panel depicts the components of action space,
 and the bottom panel the components of velocity.
 \vspace{-10pt}
 }%
\label{fig:DynamicalDist}
\end{figure*}

For a given star, the PDF of any other dynamical quantity normally calculated using 6D position-velocity can be found using the inferred \vlos{} PDF.
Whilst this can be found analytically, in practice, we find it is easier to sample the stars' \vlos{} PDFs numerically.
For each star, we sample  \vlos{} points on a linear space,
which we use to compute the 6D position-velocity and calculate dynamical quantities such as IoM.
At each of these \vlos{} points, the star's \vlos{} PDF defines a weight, normalised so that across the \vlos{} points for a given star, the weights sum to one.
Across all stars, \vlos{} distributions can then be marginalised over by using these weighted contributions to distributions.
By instead using the PDFs of individual subgroups, whilst still normalising by total, the dynamical distributions of the subgroup can be found.

This method is applied to recover the 1D dynamical distributions of the total \Halo{} and \Disc{} in Fig.~\ref{fig:DynamicalDist}.
The action distributions and \vlos{} are all in excellent agreement with the original distributions.
The velocity components are recovered reasonably well, but the halo distribution shows some slight deviations from the true distribution.
Most notably, in the distribution of $v_R$, the true \Halo{} shows distinct asymmetry, with slightly more stars falling towards the Galaxy's centre.
This asymmetry could possibly be the influence of the bar \citep{monari19SignaturesResonancesLarge} or suggest incomplete phase-mixing.
The \Halo{} stars are not exactly uniformly distributed in angle-space, as our methodology assumes, and so this minor feature is not recovered.
There is a small excess of stars predicted at near $L_{z}\approx0$, but we find it is not significant enough to affect other results.

\begin{figure}
 \includegraphics[width=\columnwidth]{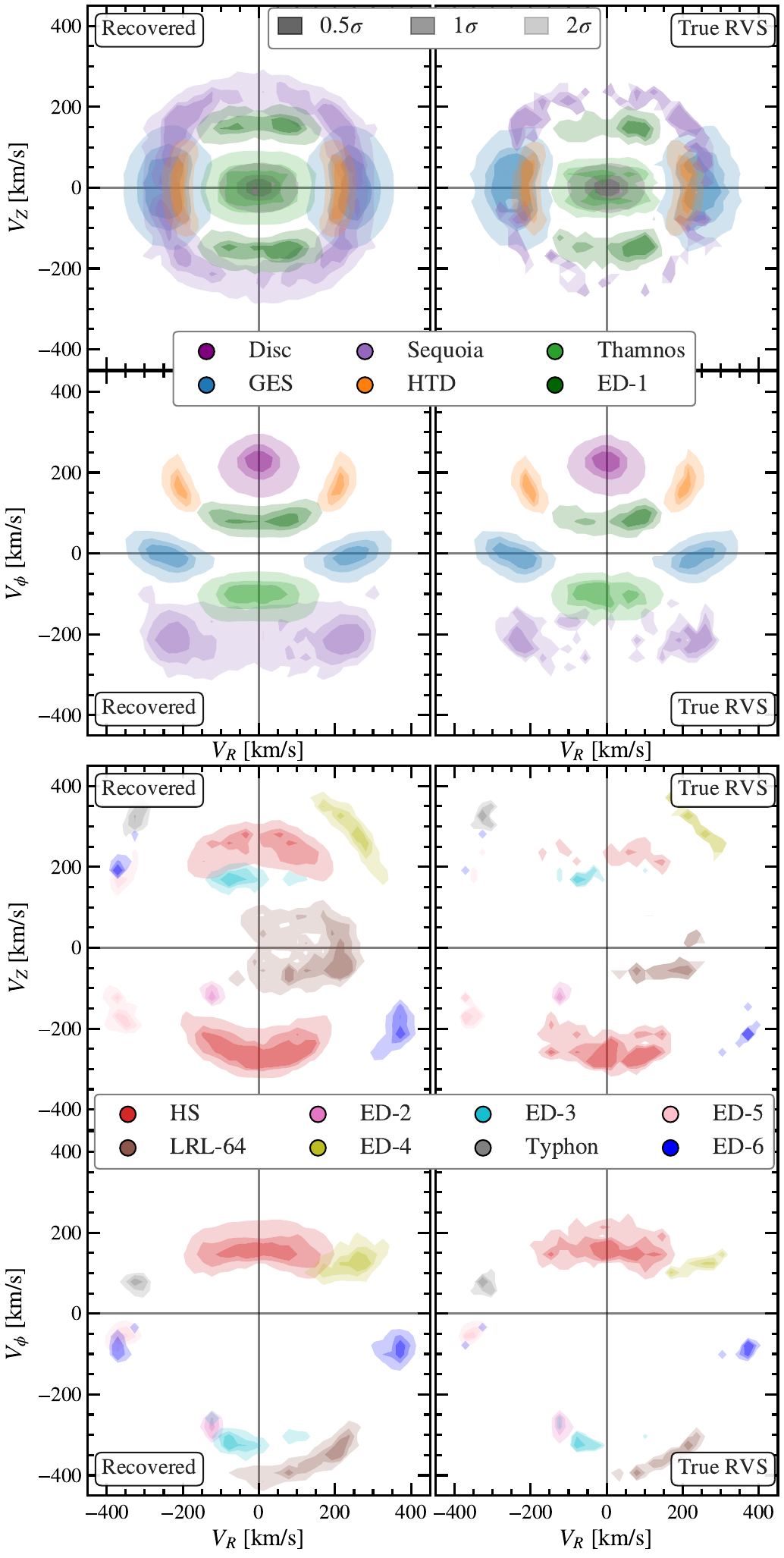}
 \vspace{-12pt}
 \caption{ Cylindrical velocity distributions of the different groups,
 as recovered by our methodology (lefthand side) against the true distributions (righthand side).
 Different stellar groups are represented by different colours,
 with the opacity of the colour representing the $\left(0.5\sigma,1\sigma,2\sigma\right)$ levels
 (equivalently the $38\%$, $64\%$, and $95\%$ regions).
 The groups are split into two separate subpanels,
 with the majority of phase-mixed groups in the top set and phase-clumped groups,
 with the addition of the Helmi Streams to allow the groups to be visually distinguished,
 in the bottom set.
 \vspace{-10pt}
 }%
\label{fig:Vel2D}
\end{figure}

This technique can also recover higher dimensional distributions
and dynamical properties of individual \Halo{} substructures, as shown in Fig.~\ref{fig:Vel2D}.
The general velocity structures of our groups are recovered very well, with additional smoothing.
The smaller asymmetries and overdensities within the velocity distributions are not recovered,
indicating a more complex phase structure in angle-space than our assumed uniform distributions.
For the phase-clumped groups in the bottom panel (excluding the phase-mixed \HS),
our additional step of explicitly selecting phase-clumped groups in velocity space ensures that clear velocity structures are well recovered.
The velocities recovered without this additional step can be seen in App.~\ref{Appendix:NoPhase}.

\section{Predicting membership probability}\label{Sec:Probs}

\begin{figure*}
 \includegraphics[width=0.98\textwidth]{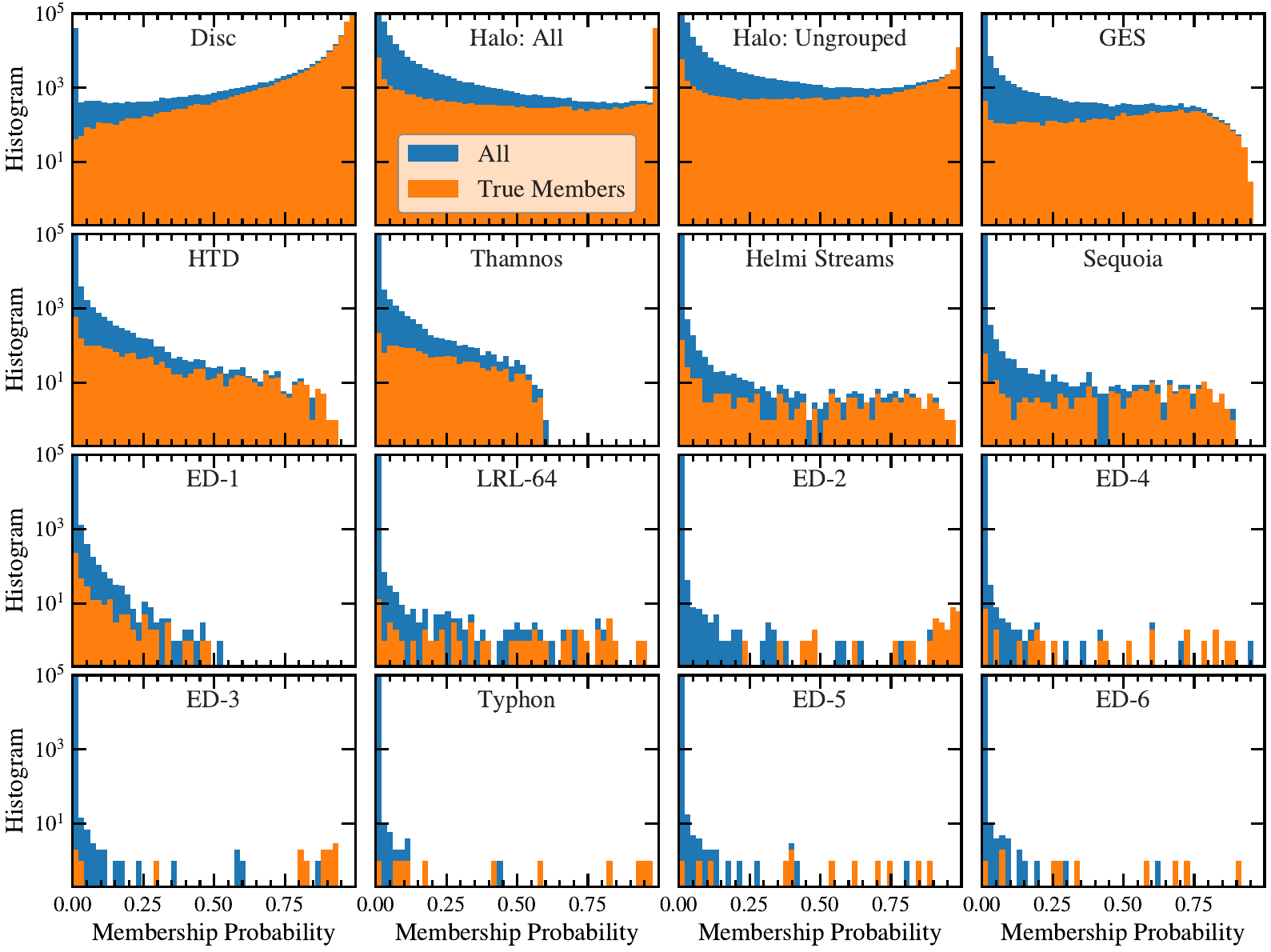}
 \vspace{-8pt}
 \caption{ 
 Distributions of membership probabilities of our individual groups (separate panels),
 calculated with our methodology.
 The distribution of the membership probabilities of every star is plotted in blue,
 and the true group members in orange.
 The first two panels depict the \Disc{} and the total \Halo.
 The total \Halo{} is decomposed into individual structures and \Ungrouped.
 By construction, a star's probabilities of belonging to the \Disc{} or the total \Halo{} sum to one,
 and the total \Halo{} membership is then decomposed into belonging to structures or \Ungrouped.
 \vspace{-10pt}
 }
\label{fig:6D_Probs}
\end{figure*}

An advantage of our methodology is that it provides the membership probabilities of a star belonging to the \Disc, \Halo,
or a specific substructure within the \Halo{}. 
These probabilities can then be used to select a sample of stars belonging to a desired substructure.
We now explore the statistical accuracy of our membership probabilities
and how effectively they can be used to select high-quality samples of likely members.

The distribution of the probabilities for each group can be seen in Fig.~\ref{fig:6D_Probs}.
The exact membership probabilities for an individual star have a complex dependency on the exact path taken through action space.
However, there are a few general trends.
Consider the probability distributions of \Disc{} and total \Halo{} stars,
whose probabilities, by definition, sum to one.
Most stars' membership probabilities between these two groups are very polarised.
The \Disc{} has overwhelming population numbers and density, 
and as a result, if a star can possibly have a \Disc{} orbit for a given range of \vlos,
it is very likely to be a \Disc{} star.
If the star cannot be on a \Disc{} orbit for any \vlos{}, it must be a \Halo{} star.
There are proportionally very few \Disc{} stars with a low probability of \Disc{} membership,
but a population of \Halo{} stars is mistaken for \Disc.

Within the \Halo,
the different substructures have very different member probability distributions.
These distributions depend on the structure's density and location in action space.
A defining feature is the lowest energy point of the action path or \Emin.
The lower a 5D stars' \Emin, the deeper the \vlos{} path goes into action space,
and the greater the range of possible \vlos{} and orbits.
In general, this correlates with larger uncertainties \sPDF{} and membership probabilities are split between different groups.

Consider \Thamnos{} and \EDone{} groups at a low-energy position in the space.
Both groups have few stars with membership probabilities greater than $50\%$,
as any star that could be a \Thamnos{} or \EDone{} member could also have an orbit at higher energy
and belong to another substructure or the \Ungrouped.
The stars' path in action space will traverse these structures,
which translates directly into alternative membership probabilities.

The smallest phase-clumped structures, such as \EDtwo, are very dense in phase space.
Few stars cross their region of phase space, but those that do are likely to be higher probability members
unless the action path also traverses an even denser structure, such as the \Disc.
On the other hand,
The \Ungrouped{} has no high-density region in action space but is instead spread over a large portion of the space,
including higher energy areas.
All stars'  \vlos{} path in action space will cross a section of this high-action space before the orbit becomes unbound.
As a result, all stars have a nonzero probability of belonging to the \Ungrouped.

\subsection{Expected populations}

\begin{figure}
 \includegraphics[width=\columnwidth]{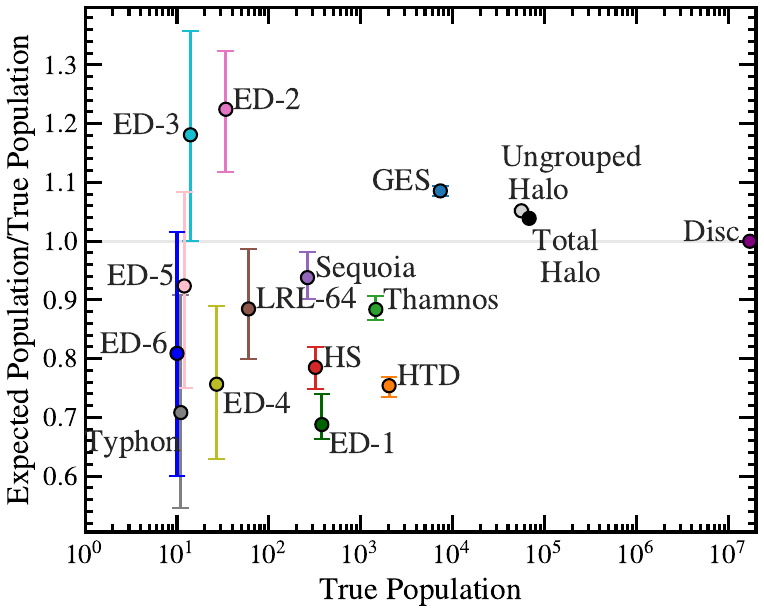}
 \vspace{-14pt}
 \caption{Expected populations of the stellar groups of our 6D RVS sample,
 calculated using membership probabilities assigned by our method,
 compared to the true population (based on \protect{\citetalias{dodd23GaiaDR3View}} groups).
 The error bars are estimated by redrawing the populations based on the membership probabilities of the stars.
 \vspace{-15pt}
 }%
\label{fig:6D_Pops}
\end{figure}

The statistically expected population number of a group in the sample can be calculated 
by considering the membership probabilities across the entire sample of stars:
\begin{equation}
 \left<N_g\right> = \sum_{i\in\mathrm{Stars}}p_{g}^{i},
\end{equation}
where $p_{g}^{i}$ is the probability that star $i$ belongs to group $g$.
The uncertainties in this estimate can be calculated by randomly drawing a group population based on the probabilities many times (we use $10,000$),
then taking the median and $16-84\%$ percentile ranges.
The accuracy of this method can be studied by comparing the estimated population to the true population (see Fig.~\ref{fig:6D_Pops}).

The ratio between the \Halo{} and the \Disc{} is approximately correct,
with $99.6\%$ of stars expected to belong to the \Disc,
with the total \Halo{} overestimated by only $\sim4\%$.
The population sizes of the substructures are typically recovered within the expected statistical uncertainties.
Overall, our groups' expected populations are within $30\%$ of the true population,
with a slight bias to underestimate smaller groups.
The smaller groups have more significant fractional uncertainties due to poor sampling and a larger Poisson effect.
Even groups without many high-probability members, such as \Thamnos{} and \EDone, 
have approximately the correct expected population due to having many low-probability members.

\subsection{Purity, completeness and selecting a sample}\label{Subsec:Purity}

\begin{figure*}[hbtp!]
 \includegraphics[width=\textwidth]{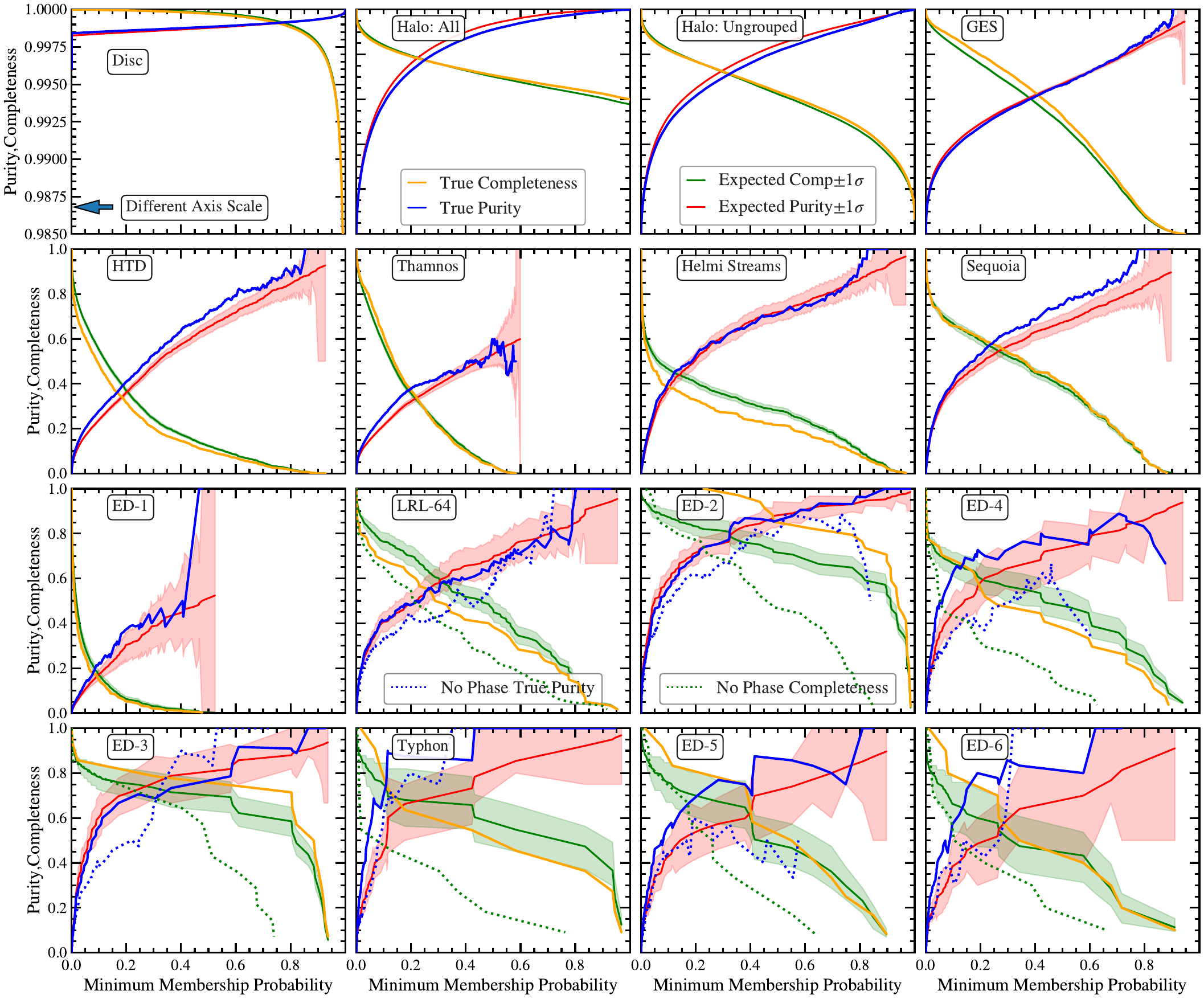}
\vspace{-12pt}
 \caption{ 
 \RefChange{Plots depicting effects of increasing the minimum membership probability upon the purity and completeness of the selected sample of a group.}
 For each group (panel), increasing the probability cut on membership probability (x-axis) increases the purity of the selected group,
 but decreases the completeness (y-axis).
 We plot the true purity and completeness,
 and the expected purity and completeness calculated from the probability distribution.
 For groups that show evidence of being phase-clumped, we also plot the purity and completeness calculated without the additional phase considerations
 (see Sec.~\ref{SubSec:phase-mixed}).
\vspace{-10pt}
 }%
\label{fig:6D_PurityComp}
\end{figure*}

A sample of stars likely belonging to a group can be selected with a cut in minimum membership probability, $\pmin$
(see Fig~\ref{fig:6D_Probs}).
The quality of the selected sample for a given $\pmin$ can be quantified by defining the purity and completeness as:
\begin{align}\label{Eq:Scoring}
 P_g\left(\pmin\right) &= N_{g}^{\cap}\left(\pmin\right) / N_{g}\left(\pmin\right)\\
  C_g\left(\pmin\right) &= N_{g}^{\cap}\left(\pmin\right) / N_{\mathrm{g}},
\end{align}
where $N_\mathrm{g}$ is the true population of the group in the total sample,
$N_{g}\left(\pmin\right)$ is the population above the \pmin{} probability cut,
and $N^{\cap}_{g}\left(\pmin\right)$ is the true population above the \pmin{} probability cut.

The effects of choosing different probability cuts are explored in Fig.~\ref{fig:6D_PurityComp}.
Increasing the \pmin{} of a sample increases the purity but reduces the completeness and overall sample size.
This behaviour at a given \pmin{} depends on the distribution of membership probabilities for the selected group (Fig.~\ref{fig:6D_Probs}).
Groups with many correct high-probability members can be recovered well, with high purity and completeness.

For an actual 5D sample without observed \vlos{}, the true purity and completeness of a selection will be unknown,
and they cannot be used to guide an appropriate selection of \pmin.
Instead, the expected purity and expected completeness of a given sample can be calculated using the expected populations at a given \pmin{} cut 
instead of the true populations:
\begin{equation}\label{Eq:Numbs}
 \left<N_{g}\right>\left(\pmin\right) = \sum_{i\in \left[p_{g}^{i} > p_{\mathrm{cut}}\right]}p_{g}^{i}.
\end{equation}
The uncertainties in these values can be found by stochastically drawing samples based on the membership probabilities,
similar to those found for the total expected populations.
These expected values are plotted as shaded regions in Fig.~\ref{fig:6D_PurityComp}.
Generally, the expected purity matches the true purity within the expected scatter,
suggesting that the purity and completeness of a given sample can be predicted.

\begin{figure*}
 \includegraphics[width=\textwidth]{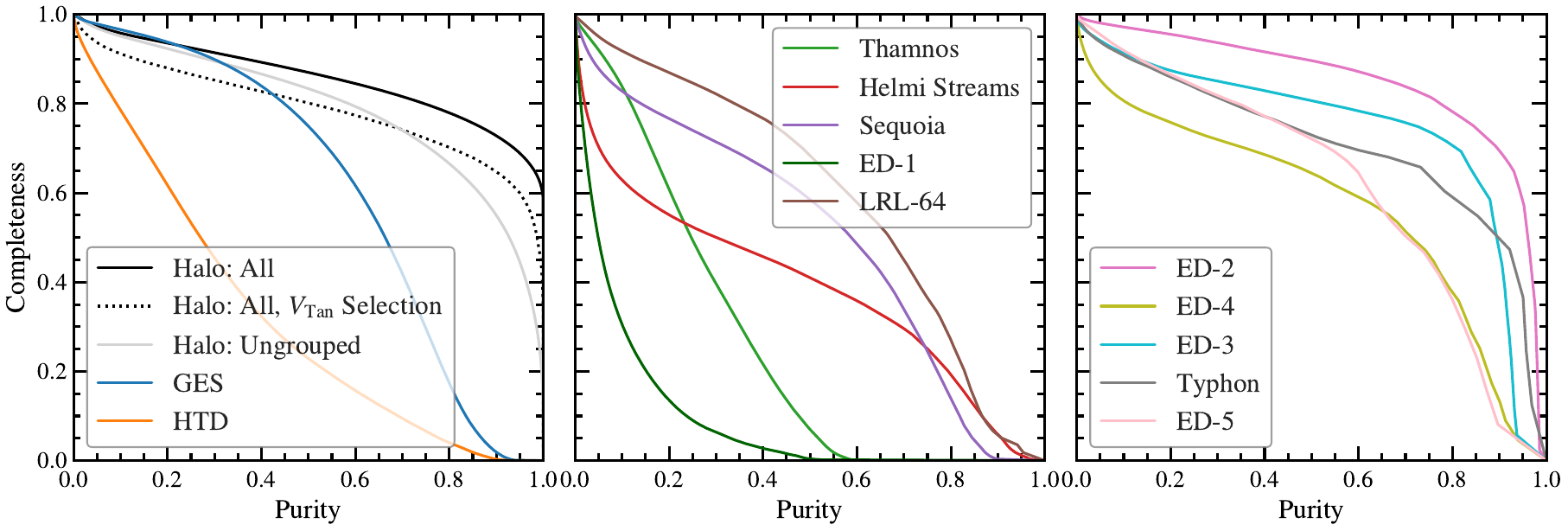}
 \vspace{-12pt}
 \caption{Curves describing the effective recovery for our different groupings,
 using the expected purity and completeness of a given probability cutoff.
 These curves describe the inherent trade-off of choosing a sample of stars;
 the higher the purity, the lower the completeness.
 Better recovered structures have a higher completeness for a given purity.
 In the first panel, we additionally plot the recovery curve from making a transverse sky velocity cut (see text for details).
 }\label{fig:tradeoff}
\end{figure*}

When selecting a sample of 5D stars, the ideal choice of \pmin{} depends on the intended analysis,
which will typically require a certain purity or number of stars.
The trade-off between expected purity and completeness can be seen directly by considering the purity
of a sample at a given completeness, parameterised by the probability cut $\pmin$ as a recovery curve, plotted in Fig.~\ref{fig:tradeoff}.
These curves concisely describe how well a given group can be recovered,
with the best-recovered groups giving higher completeness for a given purity.
For the smaller groups (rightmost panel), a purity of $80\%$ can be taken while remaining above $50\%$ complete.
For some poorly recovered groups, such as \EDone, \Thamnos, and the \HTD, any sizeable sample taken will inherently be impure.

\section{On comparisons and selections}\label{Sec:Comparsion}

\subsection{Comparison to Naik24}\label{SubSec:Naik24}
\begin{figure}
 \includegraphics[width=\columnwidth]{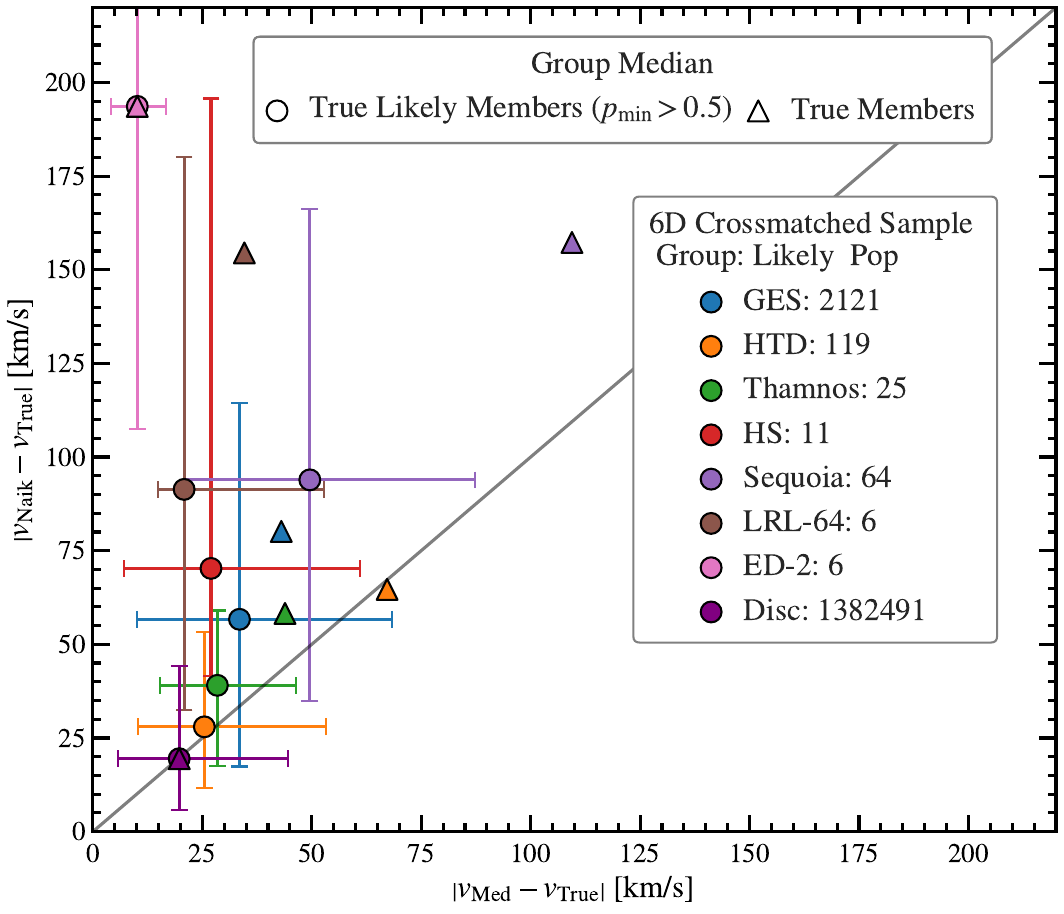}
 \vspace{-12pt}
 \caption{Comparisons between the differences between the true \vlos{}  and the predictions of our methodology and 
 {\protect{\citetalias{naik24MissingRadialVelocities}}},
 with samples decomposed into different halo substructures.
 We include all stars from the true groups, plotted with triangles,
 and true likely members with membership probability greater than $50\%$.
 Only groups with more than 4 likely members are plotted.
 The points and error bars represent the median and the $16\%$ and $84\%$ percentiles of the distribution, respectively, for the likely groups.
 These results use the crossmatched 6D sample described in the main text.
 \vspace{-10pt}
 }\label{fig:CompareErr_Substructure}
\end{figure}

Arguably, the closest current work to our predictions of \vlos{} for individual 5D stars is \citetalias{naik24MissingRadialVelocities}.
This work used machine learning on inputs of the 5D observable coordinates,
applied to a large sample of stars spanning much greater distances than our own $2.5\kpc$ limit.
In theory, the method of \citetalias{naik24MissingRadialVelocities} does not make assumptions of phase-mixing,
axisymmetry, or the Galactic potential and can predict disequilibrium structure.
They successfully recover \vlos{} with a typical uncertainty of $25-30\kms$
and demonstrate that their PDFs are statistically consistent with a value of $\Delta_{\%}=1.4\%$.
Since they do not explicitly distinguish between \Halo{} and \Disc{} members, their results can be expected to be dominated by \Disc{} stars.
Their uncertainties are very similar to our own predictions for \Disc{} stars, which, on average,
are slightly smaller than the velocity dispersion of the thin Disc.

To compare \vlos{} predictions, we use our 6D Crossmatched sample, containing approximately $30$ thousand \Halo{} stars.
Using the provided median \vlos{} estimate of \citetalias{naik24MissingRadialVelocities}, denoted $V_{\mathrm{Naik}}$,
we show the distribution of the differences $\Delta_{v}^{\mathrm{Naik}}= \left|V_{\mathrm{Naik} - V_{\mathrm{True}}}\right|$  for the different subgroups
with the red curve in the bottom panel of Fig.~\ref{fig:Verror}.
We see similar distributions of the differences $\Delta_{v}$ in the \Disc{} and \Halo{} samples.
However, the stars belonging to \Halo{} substructures do not show smaller differences $\Delta_{v}$ 
compared to the total \Halo{}, as we find with our velocity estimates.
This could suggest that not all individual substructures have been effectively modelled.

To test this more explicitly, the \vlos{} prediction differences $\Delta_{v}$  of our \vmed{} results
and those of \citetalias{naik24MissingRadialVelocities}
are directly compared for members of different structures in Fig.~\ref{fig:CompareErr_Substructure}.
Generally, for larger structures, such as the \Disc{}, \GES, \Thamnos{}, and the \HTD{}, we see very comparable results.
However, for smaller structures, our  prediction differences $\Delta_{v}$ are significantly smaller
than those of \citetalias{naik24MissingRadialVelocities},
with the most significant change for \EDtwo.
As a structure, \EDtwo{} has few members spread out very thinly over the observable space,
and so it is difficult for methods like \citetalias{naik24MissingRadialVelocities} to recover,
unlike our IoM-based methodology.

When the stars are further restricted to those that we find likely to belong to a structure ($\pmin>50\%$), we see
improvement in both estimates.
This improvement is expected for our own results, as almost by definition, the true stars that we find likely members have been well recovered.
The fact that the results of \citetalias{naik24MissingRadialVelocities} also improve
suggests that some common areas of the 5D parameter space are better recovered.

\subsection{Comparison to Mikkola23}\label{SubSec:Mikkola23}

The work of \citetalias{mikkola23NewStellarVelocity} models the local (within $3\kpc$) velocity distribution of a 5D \Gaia{} sample. 
They split their sample using a $\vtan>200\kms$ cut to divide the Disc and Halo,
with a colour-magnitude diagram selection to focus on the Halo blue sequence \citep{gaiacollaboration18GaiaDataRelease_HR}.
In the Disc, their methodology successfully finds velocity substructure, including new likely structures within the 5D dataset.
In our work, we make no distinction between moving groups of the \Disc, instead focusing solely on the \Halo.

In the Halo, their predicted velocity distributions show the signature of several prominent accreted structures,
which also appear in their recovered action space distributions.
By finding the total density within these velocity features, they estimate the population of substructures of the 5D sample.
\RefChange{In theory, the methodology can also be extended to attribute individual stars to particular structures
\citep[e.g., ][]{cronstedt23FindingStarsThat} but we cannot currently directly compare predictions for single stars.}

\subsection{Selecting a halo sample with vtan}\label{SubSec:Vtan}

\begin{figure}
 \includegraphics[width=\columnwidth]{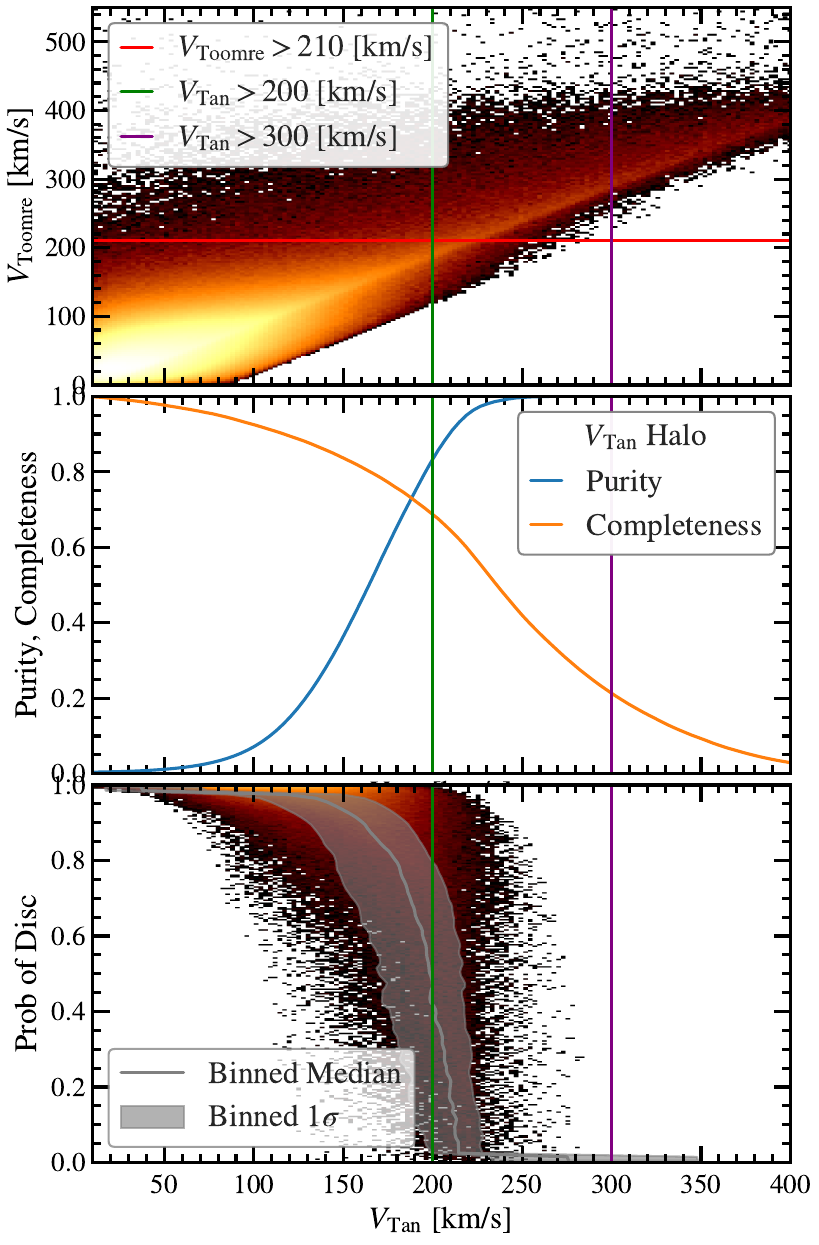}
 \vspace{-12pt}
 \caption{Plots of the dependence of \Halo{} and \Disc{} properties on \vtan, the on sky velocity.
 The top panel shows the distribution of \vtan{} against \vtoomre,
 where the horizontal red line depicts the \vtoomre{} cut used to kinematically define our \Halo{} and \Disc{} Samples.
 The middle panel shows the purity and completeness of the sample;
 for the common $\vtan>200\kms$ halo selection, the sample is $83\%$ pure and $69\%$ complete.
 The bottom panel shows the probability of the star belonging to the Disc,
 as inferred by our method.
 \vspace{-10pt}
 }
\label{fig:VtanPlot}
\end{figure}

It is common practice to define a \Halo{} sample 5D stars using only the transverse velocity of a star across the sky,
defined:
\begin{equation}
 V_{\mathrm{Tan}} = 4.74057\,\mathrm{km}\mathrm{s}^{-1}
 \left(\frac{\mu}{\mathrm{mas}}\mathrm{yr}^{-1}\right)\left(\frac{d}{\mathrm{kpc}}\right),
\end{equation}
where $d$ is the heliocentric distance, and $\mu$ is the amplitude of proper motion,
$\mu = \sqrt{\mu_{b}^{2}+ {\mu_{l}^{2}}}$.
The relationship between \vtoomre{} and \vtan{} can be seen in the top panel of Fig.~\ref{fig:VtanPlot}.

Any star with a large \vtan{} is likely to have a large \vtoomre{} and can be assumed to be kinematically a \Halo{} star.
However, the selection on \vtan{} neglects \Halo{} stars that distinguish themselves from the \Disc{} through high \vlos{}.
The halo selection cut typically used is $V_{\mathrm{Tan}}>200\kms$ \citep{gaiacollaboration18GaiaDataRelease_HR,mikkola23NewStellarVelocity},
which selects a \vtoomre{} defined \Halo{} with $83\%$ purity and $68\%$ completeness.
A stricter cut improves purity but quickly reduces the completeness, as shown in the middle panel of Fig.~\ref{fig:VtanPlot}.
$\vtan>300\kms$ gives a sample of greater than $99\%$ purity but only $21\%$ completeness.

This technique makes an interesting comparison with our methodology,
as \Halo{} stars with low \vtan{} should have orbits consistent with the \Disc{} for small \vlos{}.
As expected, there is a clear correlation between higher \vtan{} and decreasing probability of belonging to the \Disc{},
as shown in the bottom panel of Fig.~\ref{fig:VtanPlot}.
We find that our methodology offers a marginally better way of defining a halo sample,
with a slightly higher completeness for any given purity.
A similar $83\%$ purity selection has a completeness of $76\%$, 
an additional $8\%$ sample size compared to the equivalent $\vtan>200\kms$ selection.
The recovery curve from taking a varying \vtan{} cut can be seen in the first panel of Fig.~\ref{fig:tradeoff},
and is below the recovery curve of our own results.

\section{Discussion}\label{Sec:Discussion}

This work has made several implicit assumptions about the nature of our Galaxy.
Firstly, we have assumed that the Galaxy is well-modelled by an axisymmetric, static Galactic potential.
Secondly, we have assumed that the stars are completely phase-mixed
(or, for the phase-clumped stars, uniformly distributed within a reduced region of angle-space).
These assumptions have allowed us to model the local stars effectively in action-angle space.
We have found no strong biases or differences when testing our methodology with a different
galactic potential to compute the actions.

To evaluate our results, we have been comparing our inferred membership probabilities to the initial labelling of \citetalias{dodd23GaiaDR3View},
originally made in the space of energy and components of angular momentum.
Whilst these groupings are approximately consistent with regions of action space,
there is typically a difference in the membership of individual stars under $10\%$.
The groups with the largest membership difference are \Thamnos, at $30\%$, and \EDone, at $20\%$.

Like any other kinematic selection, the groups of \citetalias{dodd23GaiaDR3View} are inherently incomplete and contaminated.
Notably, the \Thamnos{} group is likely to contain large amounts of contamination.
The area of phase space that \Thamnos{} occupies at lower energy has several other overlapping groups, such as \GES{}
and splashed insitu stars \citep{belokurov20BiggestSplash}.
The ``true'' metal-poor Thamnos is probably far smaller
but would require a more comprehensive chemical selection to improve the purity of the selection.

In general, selections of \citetalias{dodd23GaiaDR3View}, based on a Mahalanobis distance of $2.2$, are likely to be conservative.
This is particularly true for \GES, whose population of selected members is proportionaly a smaller fraction of the total \Halo{}
than estimates of its stellar mass compared to the total \Halo{} \citep{helmi18MergerThatLed,lane23StellarMassGaiaSausage}.
Furthermore, as a massive accretion event, the true members of \GES{} are scattered wider in IoM space, and it is hard to make a pure or complete kinematic selection
\citep{amorisco15FeathersBifurcationsShells,koppelman20MessyMergerLarge,carrillo23CanWeReally}.

\subsection{Phase-clumped or phase-mixed}\label{SubSec:Test_PhaseMix}
\begin{figure}
 \includegraphics[width=\columnwidth]{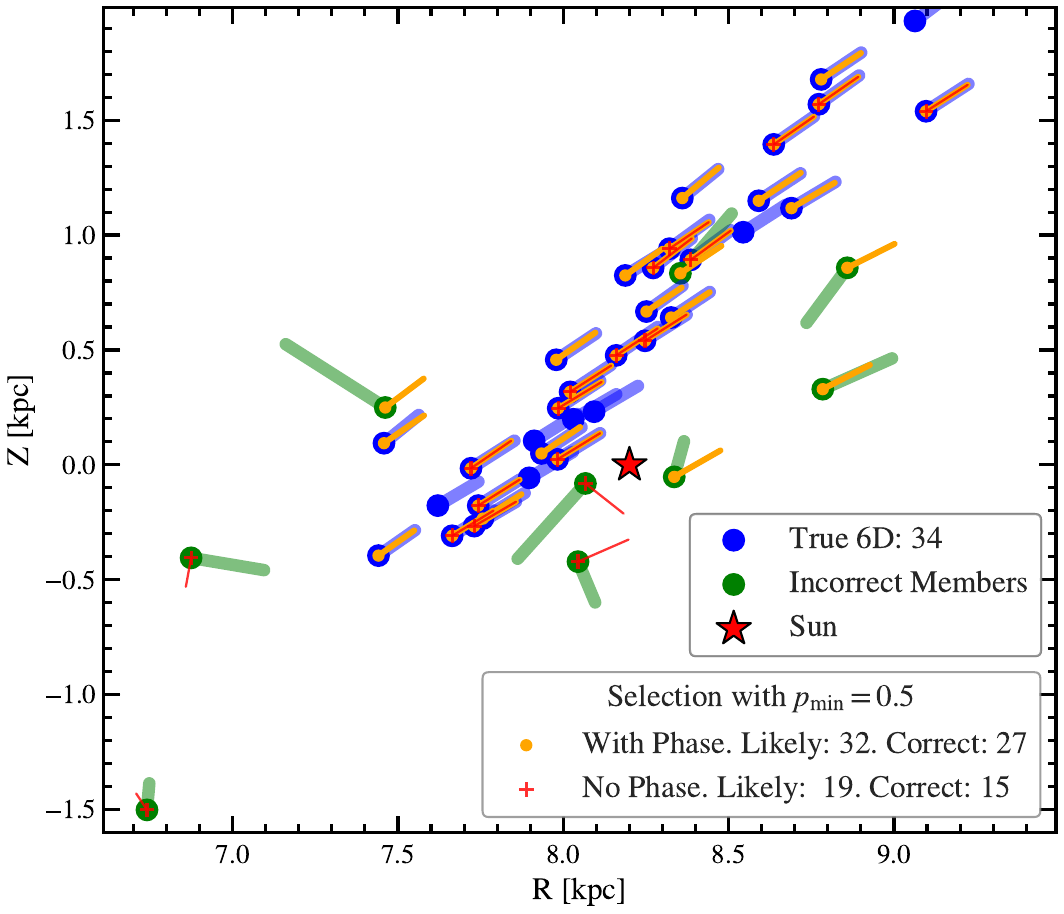}
 \vspace{-12pt}
 \caption{Stars from our ED-2 samples in R-z space, with velocities indicated as trails.
 In blue are the true known members from our 6D \Gaia{} RVS sample.
 In orange, the likely members of ED-2 inferred from 5D information, without known \vlos{}, by our method.
 The velocities are predicted from the most likely \vlos{} corresponding to the peak of the PDF.
 In red, the stars were inferred without using the additional phase information of ED-2 (described in Sec.\ref{SubSec:phase-mixed}).
 Both of these samples have been selected with a minimum probability of $\pmin=50\%$,
 corresponding to an expected purity of 86\% with phase information and 56\% when assuming the structure is phase-mixed.
 }\label{fig:ED2_Rz}
 \vspace{-10pt}
\end{figure}

In our methodology, we have modelled groups identified as phase-clumped with an additional step to account for their non-uniform distribution in angle-space.
To quantify the effect of this step, we compare our results to the methodology without this additional phase-clumped assumption.
We find that the purity and completeness of the groups modelled as phase-clumped are higher than when  modelled as phase-mixed
(see Fig.~\ref{fig:6D_PurityComp}).
The phase-clumped modelling increases the membership probability of true members and reduces the number of contaminants.

This relative improvement is most evident when considering the most phase-clumped structure, \EDtwo. 
The true 6D Gaia RVS members form a neat stream travelling through the solar vicinity (see Fig.~\ref{fig:ED2_Rz}).
The likely samples (using $\pmin=50\%$) found assuming the structure is phase-clumped correctly recovers $27$ out of $34$ true members, with 5 contaminants,
whilst the phase-mixed sample recovers only $15$ likely members, with 4 contaminants.
The phase-mixed sample contains a few obvious outliers,
stars with likely velocities (defined using \vmax) that are misaligned with the stream.
The contaminants of the phase-clumped sample seem to mostly lie outside the thin region of $\left(R, Z\right)$ space that \EDtwo{} inhabits,
suggesting that an additional selection in physical space could improve the purity of the membership.

However, it could be argued that this additional step risks overfitting these phase-clumped groups.
It is possible that these structures are poorly sampled,
and with a larger sample of stars, additional members could be found in different areas of velocity space.
These new stars would be missed by our phase-clumped modelling's strict velocity selection but would, in principle,
be detectable with phase-mixed modelling (see App.~\ref{Appendix:NoPhase}).
Furthermore, stars on resonant orbits will not be well modelled by this methodology due to their distribution in angle-space.
In our following paper, we will apply our method to a 5D \Gaia{} sample,
assuming the groups are phase-clumped to find high probability members.
We will also apply our method assuming the groups are phase-mixed and search for new features in the velocity space.

\subsection{Improving the method}

Our methodology could be developed further in several ways.
First, we have restricted ourselves to a sample within $2.5\,\mathrm{kpc}$ to ensure a high-quality,
well-sampled representation of the phase-space distribution.
If we could extend our methodology to greater distances, we would have a greater number of stars and additional structures not seen 
within the local solar neighbourhood, for example, the Sagittarius stream.
This extension would require a sufficiently large 6D sample of stars at larger distances to model the phase space effectively.
Alternatively, we could model the action distributions of the Disc and Halo using analytic distributions
\citep{li22ModellingStellarHalo,binney24ChemodynamicalModelsOur}.
These distributions would allow us to find the likely Disc and Halo membership probability for 5D stars at any distance
and not suffer from poor sampling in high-energy regions.
However, the Halo is arguably composed entirely of substructure \citep{naidu20EvidenceH3Survey},
the majority material of \GES.
Any functional distribution would be imperfect and could easily dominate our smaller substructures
or cause biases in poorly fit regions of the phase space.
Instead, we have chosen to remain as close to the data as possible.

In this work, we have only considered the kinematic information of 5D stars.
As demonstrated by \citetalias{mikkola23NewStellarVelocity}, 
the red or blue sequence has very different dynamical distributions,
which could allow stronger inference on the star's likely \vlos.
Furthermore, information from the colour-magnitude diagram could help distinguish between likely members or outliers for some simple structures.
On the other hand, as our method is purely dynamical, this additional information is unconstrained,
allowing stellar properties such as the position in the colour-magnitude diagram to verify the membership of structures independently
and properties of the group to be inferred without bias.

\section{Conclusion}%
\label{Sec:Conclusion}
In this paper, we have developed a technique to infer the likely \vlos{} and dynamics of 5D stars within $2.5\,\mathrm{kpc}$.
This allows us to derive the membership of the stars to the \Disc{} and \Halo, as well as substructures that have been characterised statistically
in the \Halo{} by \citetalias{dodd23GaiaDR3View}.
Our methodology uses a sample of 6D \Gaia{} RVS stars to estimate the density of action-angle space, 
correcting for selection effects of the \Gaia{} RVS, a distance cut, and quality cuts.
By assuming that most structures are phase-mixed, the 6D action-angle space can be reduced to 3D action space.
For phase-clumped structures,
an additional refinement is possible by assuming their stars are distributed within a restricted region of the observable angle-space.
This action-angle density is directly related to the likely \vlos{} 5D star, allowing a PDF to be inferred,
which can then be decomposed into contributions from individual groups.

We have demonstrated that our inferred \vlos{} PDFs are statistically accurate, unbiased,
and with predicted uncertainties reflecting the true differences between our \vlos{} estimates and the true values.
Overall, \Disc{} stars are far easier to identify and constrain than stars belonging to the general stellar \Halo.
However, the \vlos{} of some \Halo{} stars belonging to compact Halo substructures can be well constrained.
By marginalising over the inferred \vlos{} PDFs, dynamical information of the total sample and
individual substructures can be reliably recovered.

We find that the inferred membership probabilities of our stars are statistically consistent with our initial labelling.
We have demonstrated that these probabilities can be used to select samples of the likely members of substructures,
with the purity and completeness reliably estimated.
The trade-off between high purity and low completeness differs for every group,
and samples should be selected depending on the specific use case.
Some substructures can be recovered well, with high expected purity and completeness,
whilst others are difficult to recover due to their position in action space.
Additionally, these probabilities can be used to predict the population size of a given substructure,
typically within $30\%$ of the true value.

In comparison to a \vtan{} based selection, we find that our methodology offers a marginally better way of defining a \Halo{} sample.
For the \Disc{} and general \Halo{}, our \vlos{} predictions have similar differences to the true \vlos{} as the results of \citetalias{naik24MissingRadialVelocities}.
However, for \Halo{} substructures, we find evidence that our methodology performs better, predicting \vlos{} with greater accuracy and precision.

Whilst there is no substitute for an observed \vlos, this paper has demonstrated that
5D stars can be used for dynamical analysis.
In a following paper (in prep), we will apply our method to a local 5D \Gaia{} sample containing over 100 million stars 
to better characterise the stellar populations of known substructures.

\section*{Acknowledgements}
We acknowledge financial support from a Spinoza prize to AH. 
This work has made use of data from the European Space Agency (ESA) mission \Gaia{} (\url{https://www.cosmos.esa.int/gaia}), processed by the \Gaia{} Data Processing and Analysis Consortium (DPAC, \url{https://www.cosmos.esa.int/web/gaia/dpac/consortium}). Funding for the DPAC has been provided by national institutions, in particular, the institutions participating in the \Gaia{} Multilateral Agreement.

This work used Python and Rust programming languages.
The analysis has benefited from the use of the following packages:
vaex \citep{breddels18VaexBigData}, \textsc{AGAMA} \citep{vasiliev19AGAMAActionbasedGalaxy},
NumPy \citep{vanderwalt11NumPyArrayStructure},
matplotlib \citep{hunter07Matplotlib2DGraphics}
and jupyter notebooks \citep{kluyver16JupyterNotebooksPublishing}.

TC thanks JSW for their moral support.




\bibliographystyle{aa}
\bibliography{references.bib}



\begin{appendix}
\section{Selection function calculation}%
\label{Appendix:SF}

As discussed in Sec.~\ref{SubSec:SFCorrection}, we want to estimate the expected fraction of time \tJ,
or equivalently the expected fraction of angle-space volume,
that a typical halo population on an orbit \J{} is observable within.
Our method for calculating this is to find points upon the orbit \J{},
place stars from a typical \Halo{} population at these positions, and then try to observe them.
In practice, we calculate this process into three steps:

\begin{enumerate}
 \item Calculate the probability of seeing a given star at a single point.
 \item Calculate the expected fraction of a Halo population observed at a single point.
 \item Calculate the expected fraction of a Halo population observed on an orbit \J.
\end{enumerate}

\subsection*{1. Observational probability of a single star at a single point}
First, we must be able to estimate the probability that a given star,
with known stellar properties and position is in our data sample.
When a star of intrinsic absolute magnitude and colour 
is observed, its apparent magnitude and colour are affected by extinction,
depending on the local dust distribution and its physical position relative to the Sun.
We use the dust map of \citet{lallement22UpdatedGaia2MASS3D}
and the coefficients of \citet{gaiacollaboration18GaiaDataRelease_HR} to apply reddening and de-reddening corrections.

The \Gaia{} selection function is complex and the subject of various studies \citep{boubert22SelectionFunctionToolbox,castro-ginard23EstimatingSelectionFunction}.
In \citet{castro-ginard23EstimatingSelectionFunction},
the RVS selection function has been estimated to depend on the sky position and apparent magnitude and colour of the star
$\left(l,b,G,G-G_{RP}\right)$.
Given these observables,
these authors provide a method to estimate the probability that the star would be included in the \Gaia{} RVS catalogue.
Numerically, the RVS function information is defined with a physical grid of healpix bins and distances,
followed by a grid of colours and magnitudes.

We model the quality cuts on relative parallax error and radial velocity error
by estimating these uncertainties with the \textsc{PyGaia}\footnote{https://github.com/agabrown/PyGaia} software package.
This software provides a model estimating the average uncertainty of observed quantities given the apparent magnitude and colour.
In practice, these modelled uncertainty cuts make a slight difference on top of the modelled \Gaia{} RVS selection function,
with the parallax cut affecting very few stars and positions,
and the radial velocity cut affects practically none.
The other quality cuts are believed to make very little difference and are neglected.

Taken together, along with the distance cut,
we have a prescription for calculating the probability that a star 
of absolute magnitude and colour, found at position $\left(l,b,d\right)$
will be included in our 6D dataset,
defined as $p\left(l,b,d,G,G-G_{RP}\right)$.
If the star falls outside of our distance and quality selection, the probability is zero.
Notably, these selection functions are independent of velocity, 
particularly \vlos.

\subsubsection*{2. Expected observed fraction of a halo population at a single point}

Instead of considering the probability of seeing a single star at a point in space,
we must consider the expected fraction observed of a typical Halo stellar population.
To do this, we draw a stellar population from an isochrone,
using a Chabrier \citep{chabrier03GalacticStellarSubstellar} Initial Mass Function (IMF).
We use an alpha-enhanced BaSTI isochrone \citep{pietrinferni21UpdatedBaSTIStellar}
with a metallicity of $\left[\mathrm{Fe}/\mathrm{H}\right]=-1.1$ and an age of $11$ Gyrs.

The isochrone is defined as a set of initial masses, absolute magnitudes and colours.
For each defined point in the isochrone,
we find the fractional number of stars associated with it,
using the IMF evaluated at its initial mass and the width of the linear mass bin it defines.
For each point on our physical grid of healpix points on the sky and distances between $0-2.5\,\mathrm{kpc}$,
we evaluate the observable probability and average the magnitude-colour points weighted by the IMF value.
On our physical grid, we now have defined the expected observable fraction of the halo population,
defined as $\tau\left(l,b,d\right)$.

We similarly define our probability information
using the same physical grid limited to distances between $0$ and $2.5\kpc$.
For each bin centre, given an absolute magnitude and colour,
we redden the absolute magnitudes and colours using the dustmap to find the ``observed'' apparent values,
which are then used to evaluate the RVS selection function and quality cuts.

\subsection*{3. Expected observed fraction of a halo population on an orbit}

To find the expected observable fraction of our \Halo{} population on a given orbit \J{}, we sample points from around the orbit.
We integrate along the orbit over time, sample its position at many points,
find the expected fraction of the halo population seen at that point, and then average over the points.

Numerically, we integrate the orbit \J{} for $40\,\mathrm{Gyrs}$ and randomly select its position at $100,000$ times.
For members of phase-clumped groups with a higher dependence upon exact position, we use
$100\,\mathrm{Gyrs}$ and randomly select its position at $2,000,000$ times.
We have checked that our results are not sensitive to these choices of time resolution.

For the majority of the times sampled along the orbit, the star will not be within $2.5\kpc$ of the Sun,
and for some of the orbits, this will be extremely rare.
To improve our sampling of a nonzero observable probability, we exploit our axisymmetric assumption.
At each sampled position, we calculate the cylindrical coordinate $\left(R,z\right)$.
If the $\left(R,z\right)$ is not within $2.5\kpc$ of $\left(R=R_{\odot},z=0\right)$, then the observable fraction is evaluated as zero.
Otherwise, we calculate the angular range of $\phi$ that the star is within the solar sphere of radius $2.5\kpc$.
We rotate the star through linearly spaced $100$ of this $\phi$ range,
finding the positions in Galactic, then observable heliocentric coordinates,
and then evaluating the expected observable fraction at the point.
These can then be averaged over the $\phi$ range, including the length of the range over the entire $2\pi$ space,
giving the evaluation for a single point on the orbit.
These points can then be averaged over to give the total expected observable fraction of the halo population for the given orbit \J{}.

\section{Density estimation}%
\label{Appendix:Density}
We want to estimate a density from the set of points and weights $\left[\boldsymbol{J}_i, w_i\right]$.
First, we transform from action space  $\boldsymbol{J} = \left(J_R,J_z,J_{\phi}\right)$ 
to the transformed space denoted $\sqJ=\left(\sqrt{J_R},\sqrt{J_z},J_{\phi}\right)$. 

Calculating the Jacobian factor, the transformation is given by:
\begin{align*}\label{Eq:DensTransform}
 F\left(J_{R},J_{z},L_{z}\right) &= \frac{1}{4\sqrt{J_{R}+\delta_{J}}\sqrt{J_{z}+\delta_{J}}}\times\\
 & \rho\left(\sqrt{J_{R}+\delta_{J}}, \sqrt{J_{z}+\delta_{J}}, L_{z}\right),
\end{align*}
where $\delta_{J}=2 \uJ$ is a small value to ensure near zero action does not correspond to unrealistic density spikes,
and $\rho\left(\tilde{\boldsymbol{J}}\right)$ is the density in the transformed space.

The problem is now to model the density $\rho\left(\tilde{\boldsymbol{J}}\right)$ as a collection of points and weights
$\left[\sqJ_i, w_i\right]$.
For this, we follow a modified implementation of MBE from \citet{ferdosi11ComparisonDensityEstimation}.

\begin{equation}
\rho\left(\sqJ|\left[\sqJ_i, w_i\right]\right) =
 {\left(\sigma_{R}\sigma_{Z}\sigma_{\phi}\right)}^{-1}\sum_{i=1}^{N} w_i{\lambda_{i}}^{-3}
 K_{e}\left(\frac{1}{\lambda_i}\left|\frac{\sqJ - \sqJ_i}{\boldsymbol{\sigma}}\right|\right),
\end{equation}
where $\left(\sigma_{R},\sigma_{Z},\sigma_{\phi}\right)$ defines the window width in each dimension and
$\lambda_{i}$ the local bandwidth parameters. 
The Epanechnikov kernel $K_e$ is defined as:
\begin{equation}
 K_e\left(r\right) = \begin{cases}
 \frac{d+2}{2V_d}\left(1-r^2\right) &r^2<1\\
 0 &\text{otherwise},
 \end{cases}
\end{equation}
where $V_d$ is the spherical volume that normalises the equation such that the kernel integrates to $1$ over the space.

We estimate the values of $\sigma_{j}$ and a collection of $\lambda_{i}$ for each group individually
so that every group's density is smoothed correctly.
For each set of group points, $\sigma_{j}$ is estimated as:
\begin{equation}
\sigma_{j} = \frac{P_{100-\omega\%}\left(\tilde{J}_j\right) - P_{\omega\%}\left(\tilde{J}_j\right)}{\log\left(N\right)},
\end{equation}
where $P_{\omega\%}\left(J_{j}\right)$ corresponds to the $\omega\%$ of the quantile of the distribution of action $J_j$.
The smoothing window parameter $\omega$ influences the smoothing;
the larger the window (the smaller the $\omega\%$), the greater the smoothing.
We use $\omega=20\%$ for all groups except the \Disc{} and \Ungrouped, which have $\omega=10\%$ to ensure they are not under-smoothed.

For the grouped stellar halo structures, we find the best performance when $\lambda_i$ is set to $1$.
For the \Disc{} and the \Ungrouped{} groups,
$\lambda_i$ are estimated through a calculation of the pilot density $\hat{\rho}$,
where the weights and $\lambda_i$ are initially assumed to be equal to 1.
Then
\begin{equation}
\lambda_{i} = {\left(\frac{\hat{\rho}\left(\sqJ_{i}\right)}{g}\right)}^{-\frac{1}{3}},
\end{equation}
where g is the geometric mean of $\hat{\rho}$.
We find that our smoothing performance appears to improve with iterations of this calculation,
with convergence approximately reached within $5$ cycles.
To ensure that structures are not under or over-smoothed,
we restrict the values of $\lambda_{i}$ in a certain range for each group.
The \Disc{} values are fixed such that $0.8<\lambda_{i}<10$,
and the \Ungrouped{} values are fixed such that $2<\lambda_{i}<10$,
reflecting the challenge of not under-smoothing the \Ungrouped{} density.

\section{Disc disequillibirum}\label{App:Disc}
\begin{figure*}
 \includegraphics[width=\textwidth]{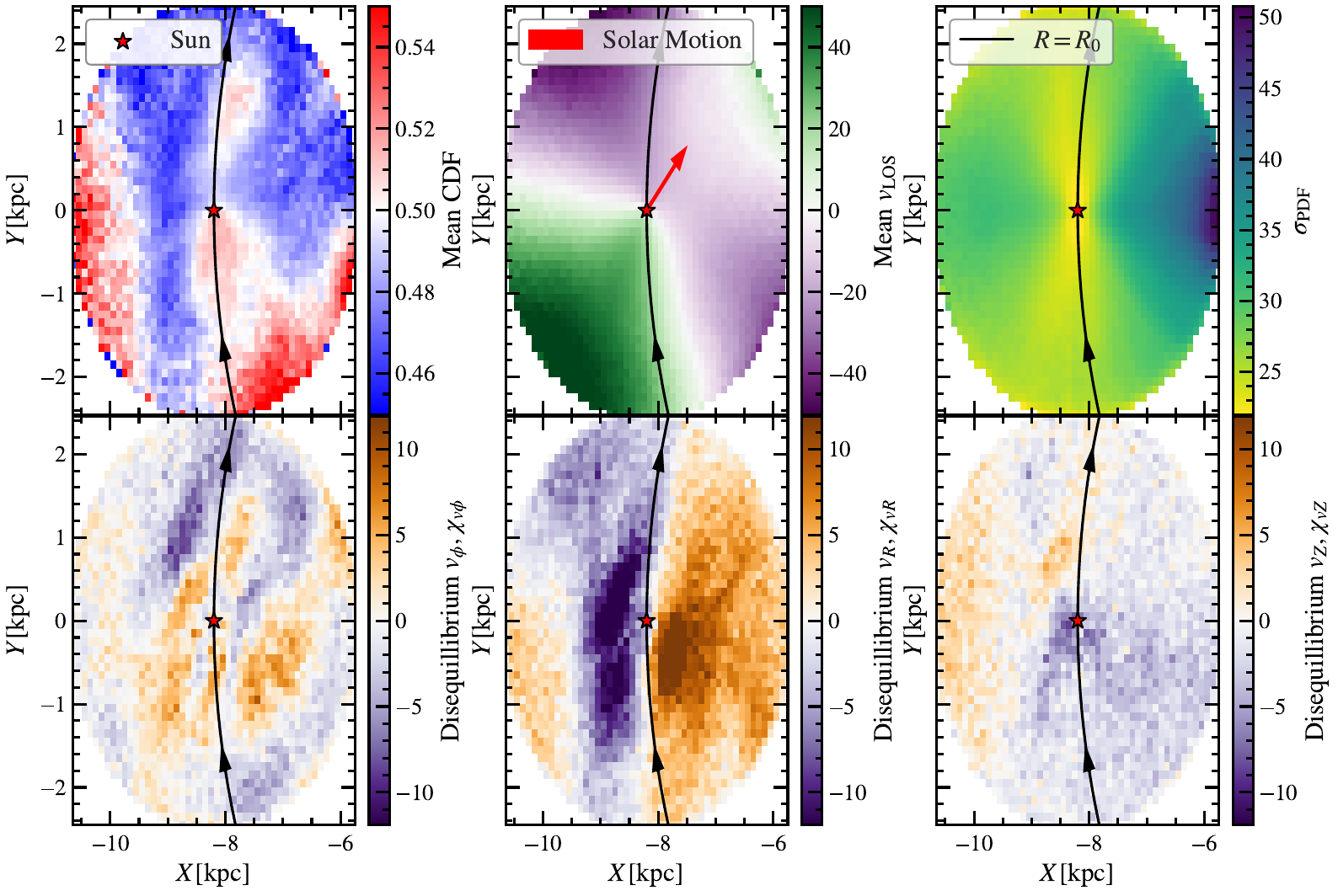}
 \vspace{-12pt}
 \caption{
 \Disc{} stars (defined kinematically by $v_{\mathrm{toomre}}<210\,\kms$) binned in galactic X and Y.
 The top panels show the mean CDF of our \vlos{} predictions (expected to be unbiased at $0.5$),
 the mean \vlos{} in the $XY$ bin due to solar motion and the expected structure of the Disc,
 and the average uncertainty in our \vlos{} prediction.
 The bottom panels show evidence of disequilibrium in the tangential, radial, and vertical
 velocity components of the stars,
 defined by the statistical significance of the deviation in the number of stars above or below the equilibrium values.
The black arc and arrows show the local standard of motion.
 }\label{fig:DiscDisEq}
\end{figure*}

In the distribution of CDF values (Fig.~\ref{fig:6D_VlosCDF}),
we saw evidence of small biases in our estimates of \vlos{} for \Disc{} stars.
To explore whether this effect is location dependent, we study our \Disc{} stars in the $XY$ plane in Fig.~\ref{fig:DiscDisEq}.
The first panel in the top row shows the mean CDF value in that bin. 
If the distribution of CDF values is unbiased and symmetric, the mean value should be 0.5,
and deviation from this indicates a bias in the true \vlos{} values in that bin.
We expect the distribution of CDF values to be uniform under any selection made,
provided that the selection is independent of the true \vlos{}.
If a subsample is biased and statistically significantly different from uniform,
it indicates that an assumption of our methodology, such as axisymmetry and phase-mixed Disc,
is being violated within that region.

We see clear structures within our mean CDF values in the $XY$ plane,
indicating areas where our PDFs are systemically over and underestimating the \vlos{}.
This structure has no clear association with the number density of the sample or the binned mean \vlos{} (middle panel).
These trends are expected from a uniform Disc distribution
and the Sun's small relative velocity within the Disc, as indicated by the red arrow
(in agreement with \citetalias{naik24MissingRadialVelocities}, Fig.~8.).
There is also no clear trend with distance.
The third panel on the top row shows the mean \sPDF{} binned in $XY$,
which has little dependence on distance and is mainly dependent on the on sky position of the star,
and it's \vtan.

Instead, we look at the dynamics within the \Disc{}
using the cylindrical components of velocity $\left(v_R,v_{\phi},v_Z\right)$
\RefChange{(using the true RVS \vlos{} values)}.
We consider the deviation from expected equilibrium values to quantify disequilibrium in velocity.
If the \Disc{} is in phase-mixed equilibrium in every bin,
we expect even numbers of stars moving up and down through the disc and in and out towards the centre of the Galaxy.
Defining $N_{+}$ and $N_{-}$ as the number of stars with a positive and negative velocity sign, respectively,
we can estimate the statistical significance of any imbalance by modelling the distribution as a binomial,
which can be approximated by a normal distribution. 
The statistical likelihood of getting $\left(N_{+},N_{-}\right)$ in units of sigma
is then $\chi$:
\begin{equation}
\chi = \frac{1}{\sqrt{N}}\left(N_{+} - N_{-}\right),
\end{equation}
where $N$ is the total number of stars in the bin.

For the tangential component $v_{\phi}$, we can define $\left(N_{+},N_{-}\right)$
as the number of stars in the bin above or below the median value of $v_{\phi}$ for stars at its given radius $R$,
using a series of radial bins.
In equilibrium, this should again be a binomial distribution,
as there is an even probability that a star is above or below the median value.
This quantity shows the deviation in the number of stars above or below the equilibrium values.

We plot these $\chi$ values for each of the velocity components in the bottom panels.
They show strong, clear evidence for dynamical disequilibrium in \Disc{} stars.
The left panel showing the tangential value has very good agreement with the local spiral arm structure,
as can be seen in \citet{poggio21GalacticSpiralStructure} Fig. 5.
The middle radial panel is consistent with \citet{vislosky24GaiaDR3Data},
and at the solar position, we are at the edge of a structure that is moving radially outward.
The final panel of vertical motion shows a slight structure at a larger radius, away from the centre.
There is more structure when comparing above and below the disc plane,
which is likely linked to the \Gaia{} phase space spiral
\citep{antoja18DynamicallyYoungPerturbed,hunt22MultiplePhaseSpirals}.

These disequilibrium structures show correlations with our areas of under and over-predicting \vlos{},
approximately matching the expected behaviour of seeing an excess in the median motion of stars.
This effect is a small bias, and we do not expect to recover the disequilibrium structure.

\section{Phase-clumped or phase-mixed}%
\label{Appendix:NoPhase}

\begin{figure}
 \includegraphics[width=\columnwidth]{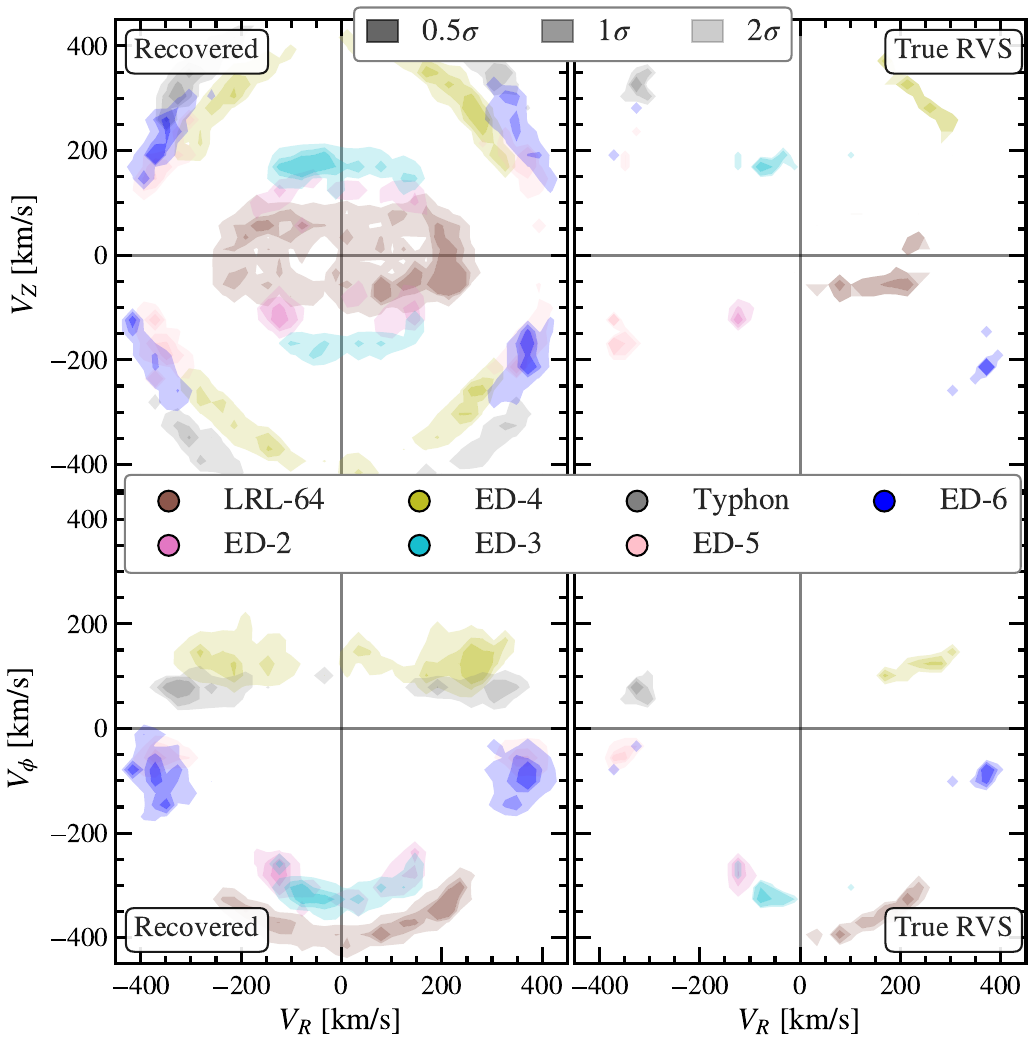}
 \vspace{-12pt}
 \caption{ 
  Cylindrical velocity distributions of the different stellar \Halo{} groups,
  as recovered by our methodology (lefthand side), against the true distributions (righthand side).
  Different groups are represented using different colours,
  with the opacity of the colours indicating $\left(0.5\sigma,1\sigma,2\sigma\right)$ levels (equivalently $38\%,64\%,95\%$).
  This figure is a comparison to Fig.~\ref{fig:Vel2D},
   where the phase-clumped nature of the selected groups are additionally modelled with an additional step.
  Here are the results of assuming all groups are perfectly phase-mixed,
   resulting in near-symmetric velocity distributions in the radial and vertical components.
 \vspace{-10pt}
 }%
\label{fig:Vel2D_NoPhase}
\end{figure}

If we assume that all groups are phase-mixed,
we find that the recovered velocity distributions are approximately evenly distributed 
between the signs of $v_Z$ and $v_R$ (see Fig.\ref{fig:Vel2D_NoPhase}).
This methodology does not model the known 6D sample as effectively but makes fewer assumptions about the nature of the stellar groups.
When modelling the true 5D \Gaia{} dataset, it is important to consider the results of both methods to check for any new additional wraps or structures.

\label{LastPage}
\end{appendix}
\end{document}